\documentclass[12pt,american]{article}
\usepackage[T1]{fontenc}
\usepackage[latin1]{inputenc}
\usepackage{geometry}
\usepackage{amsmath}
\usepackage{babel}

\makeatletter

\providecommand{\LyX}{L\kern-.1667em\lower.25em\hbox{Y}\kern-.125emX\@}

\newcommand{\lyxaddress}[1]{
  \par {\raggedright #1 
  \vspace{1.4em}
  \noindent\par}
}

\usepackage{axodraw}
\makeatother
\makeatother

\begin{document}

\title{\hfill{}{\normalsize PM/02-05}\\
Neutrino radiative decay and lepton flavor violation in supersymmetric models }

\author{M.B.Causse\thanks{
causse@lpm.univ-montp2.fr
} }

\maketitle

\lyxaddress{Groupe d'Astroparticules de Montpellier,IN2P3/CNRS,Université MontpellierII
Place E.Bataillon,CC085,34095 Montpellier Cedex 5 France}

\begin{abstract}
The radiative decay of neutrino,such as \( \overline{\nu _{\tau }}\: \longrightarrow \: \nu _{\mu }+\gamma  \)
,in the minimal supersymmetric standard model with right-handed neutrinos (MSSMRN)
and in the \( SU(5) \) grand unified theory with right-handed neutrinos(\( SU(5)RN \))model
is examined.In these models,the transition magnetic moment of neutrino is directly
proportional to the neutrino mass.For different values on SUSY parameters (compatible
with the relic density ) and for large mixing angle solution to the atmospheric
neutrino problem,the transition magnetic moment can be a few times greater than,or
comparable to,the value of the standard model (with a right-handed neutrino
singlet added).
\end{abstract}

\section{Introduction}

There is now rather convincing evidence that neutrinos have non-zero mass.This
evidence comes from the apparent oscillation of atmospheric neutrinos,solar
neutrinos and the LSND experiment\cite{4}.In the minimal standard model (SM)there
is no right neutrino that can couple to left neutrino,so the neutrinos are massless.There
are however,many possibilities to extend the minimal standard model(SM)to have
massive neutrinos.So,in the field theory description of neutrinos,by introducing
so-called Majorana mass terms\cite{1},one can split a Dirac neutrino D:\( \: \Psi =\: \Psi _{L}+\Psi _{R} \)(L,R
for left and right),into two non-degenerate Majorana neutrinos \( \chi  \)
and \( \omega  \) such as:\( \: \chi =\Psi _{L}+\Psi _{L}^{c}\: ,\: \omega =\Psi _{R}+\Psi _{R}^{c}\: where\: \chi =\chi ^{c}\: and\: \omega =\omega ^{c} \)
are self conjugate fields.Then,when both Dirac and Majorana mass terms are simultaneously
present we have: 
\[
\begin{array}{c}
L_{DM}=m_{D\: }\overline{\psi _{L}}\psi _{R}+A\: \overline{\psi _{L}^{c}}\psi _{L}+M_{R\: }\overline{\psi _{R}^{c}}\psi _{R}+h\: .\: c\\
=\frac{1}{2}\: m_{D}\: (\overline{\chi }\omega +\overline{\omega }\chi )+A\: \overline{\chi }\chi +M_{R}\: \overline{\omega }\omega \\
=\left( \overline{\chi },\overline{\omega }\right) \: \left( \begin{array}{cc}
A & \frac{1}{2}m_{D}\\
\frac{1}{2}m_{D} & M_{R}
\end{array}\right) \: \left( \begin{array}{c}
\chi \\
\omega 
\end{array}\right) .
\end{array}\]

We take \( A=0 \),then we introduce a Majorana mass for right-handed neutrino
but not for left-handed.The Dirac masses \( m_{D} \) are assumed to be comparable
to quark(lepton) masses arising from \( SU(2)_{L} \) breaking in the SM;\( M_{R} \)
reflects the scale \( \gg 1TEV \).So,it is natural to take \( m_{D}\ll M_{R} \).Then
the mass matrix \( \: \left( \begin{array}{cc}
0 & \frac{1}{2}m_{D}\\
\frac{1}{2}m_{D} & M_{R}
\end{array}\right) \:  \) can be diagonalized to give two mass eigenvalues:\( M_{1}\approx \frac{m_{D}^{2}}{M_{R}}\: ,\: M_{2}\approx M_{R} \)
corresponding to the Majorana mass eigenstates:\( \: \eta _{1}=\cos (\theta )\chi -\sin (\theta )\omega \: ,\: \eta _{2}=\sin (\theta )\chi +\cos (\theta )\omega  \)
with \( \: \tan (2\theta )\simeq \frac{m_{D}}{M_{R}}\rightarrow 0 \).So \( \eta _{1}\,  \)is
practically left-handed and very light \( (M_{1})\,  \)and \( \, \eta _{2}\,  \)
is almost right-handed and heavy \( M_{R} \).So \( \eta _{1}\,  \) is considered
as the neutrino of \( SU(2)\,  \).This is the seesaw mechanism .Thus \( \eta _{1}\,  \)
is a candidate for one of the light neutrino mass eigenstates which make up
\( \nu _{e}\, ,\, \nu _{\mu \, }and\, \nu _{\tau \, } \).So long as \( \eta _{2\, } \)
is heavy,the seesaw relation explains,without fine tuning,why a mass eigenstate
component of\( \, \nu _{e\, },\, \nu _{\mu }\,  \)or \( \nu _{\tau } \) seems
light.So the transition magnetic moment of neutrino,in the SM with a right-handed
neutrino added,is \( K_{SM}\, \cong \, 1.32\times 10^{_{-19\, }}\frac{m_{\nu }}{ev}\,  \)
(\( m_{\nu \, } \)is the neutrino mass).This value is very small comparatively
to experimental bounds.Indeed,lepton flavor conservation is an automatic consequence
of gauge invariance and the renormalizability of the minimal standard model
lagrangian.The smallness of lepton flavor violation(LFV) in charged lepton decays
\cite{2}requires large \( M_{R} \) to achieve efficient GIM.However in SUSY
models,there is a new source of flavor mixing in the mass matrices of SUSY partners
for leptons and quarks.So the magnetic moment of a massive Dirac neutrino has
been calculated in SUSY models\cite{9},\cite{11}.Unfortunately,the SUSY contribution(for
a Dirac neutrino)to the magnetic moment stays comparable to the value of the
standard model(with right-handed neutrino added).However in SUSY models with
a neutrino mass generation mechanism of the seesaw type(\cite{5},\cite{6}and\cite{10}),the
Yukawa coupling constants among the higgs doublet,lepton doublets and right-handed
neutrinos(RN)could induce large flavor mixing effects in the slepton sector.The
resulting LFV rates,for charged lepton decays,can be as large as,or even larger
than experimental upper bounds\cite{5},depending on various parameters especially
on the Majorana mass of the right-handed neutrino.In such case,the same Yukawa
coupling constants for right-handed neutrino are responsible for both the neutrino
oscillation and LFV process of charged leptons.So in this context,we study the
transition magnetic moment of the Majorana neutrino\cite{12}(with CP \emph{conserved
and} opposite eigenvalues).Thus,we calculate the radiative decay of neutrino:\( \overline{\nu _{\tau }}\: \longrightarrow \: \nu _{\mu }+\gamma \,  \).In
our calculation,we apply the multimass insertion method \cite{6}.We incorporate
the mixing of the slepton masses as well as the mixing in the chargino sector
and we consider the case of large mixing between second and third generation
suggested by the atmospheric neutrino problem.For the right-handed neutrino
sector,Yukawa couplings are assumed similar to up-type quarks ones.In our numerical
analysis,we give the transition magnetic moment for different \footnote{%
These values have been obtained with an interface that relies DarkSUSY and Suspect2
programs . This interface has be done in the PCHE gdr (degrange@poly.in2p3.fr)
}values on SUSY parameters compatible with the relic density . 

This paper is organized as follows.In section 2 we present generalities on the
lepton flavor violation (LFV) in the MSSM and in the supersymmetric \( SU(5) \)
grand unified theory with right-handed neutrino.In section 3,we give the explicit
expressions of the amplitudes associated to the chargino contributions to the
radiative decay \( \overline{\nu _{\tau }}\: \longrightarrow \: \nu _{\mu }+\gamma \,  \).Finally
in section 4,we find the numerical results on the transition magnetic moment
for different values on SUSY parameters and two right-handed neutrino masses.In
section 5 we conclude.In annex A the Feynman rules \cite{5}relevant for our
calculation are given and in annexB we show the analytic expressions for the
Feynman integrals which appear in the evaluation of the amplitudes.

\section{Generalities on the LFV in MSSMRN and \protect\( SU(5)RN\protect \) models}

\subsection{LFV in the MSSMRN model}

The MSSMRN(\cite{3}\cite{5}-\cite{6})is the simplest supersymmetric model
to explain the left-handed neutrino masses.It results from the MSSM with a neutrino
mass generation mechanism of the seesaw type.The superpotential of the MSSM
with a right-handed neutrino singlet,in the higgs and lepton sectors is given
as :

\begin{equation}
\label{superpot1}
W_{MSSMRN}\: =\: f^{ij}_{\nu }\: H_{2}\, \overline{N_{i}}\, L_{j}\: +\: f^{ij}_{e}\, H_{1}\, \overline{E_{i}}\, L_{j}\: +\: \frac{1}{2}\, M^{ij}_{\nu }\, \overline{N_{i}}\, N_{j}
\end{equation}

where L is a chiral superfield for left-handed lepton,\( \overline{N}\; and\; \overline{E}\:  \)
are for the right-handed neutrino and right-handed charged lepton.\( M_{\nu \: } \)is
a Majorana mass for right-handed neutrino.\( H_{1}\: and\: H_{2}\:  \)are for
the higgs doublet in the MSSM and i,j generation indices.In a basis where the
Yukawa matrix \( \: f^{ij}_{e}\:  \)and the Majorana mass matrix\footnote{%
after redefinition of the fields by using unitary transformations
} \( \: M_{\nu }^{ij}\:  \)are diagonalized as \( \, f_{ei}\delta ^{ij}\,  \)and
\( M_{Ri}\delta ^{ij}\,  \),neutrino Yukawa couplings \( f^{ij}_{\nu }\,  \)
are not generally diagonal.In this particular case there is LFV and then neutrino
oscillation.So neutrino Yukawa couplings can be taken as: 

\[
f^{ij}_{\nu }\, =U^{ik}\, f_{\nu k}\, V^{kj}\: \: with\: \: U\: ,\: V\: unitary\: matrices\: .\]

Therefore the Dirac masses are: 

\( m_{\nu iD}=f_{\nu i}\, v\, \frac{\sin (\beta )}{\sqrt{2}}\; \; for\; v^{2}=2\left( \left\langle h_{1}\right\rangle ^{2}+\left\langle h_{2}\right\rangle ^{2}\right) \approx \left( 246GEV\right) ^{2} \)\( and\; \tan (\beta )=\frac{\left\langle h_{2}\right\rangle }{\left\langle h_{1}\right\rangle }\; where\: \left\langle h_{1}\right\rangle \: ,\: \left\langle h_{2}\right\rangle  \)
are the higgs vacuum expectation values.

To write the neutrino mass matrix induced by the seesaw mechanism,we suppose
quark and neutrino mass hierarchy \cite{4}

\( \, m_{\nu _{\tau }}\, \gg \, m_{\nu _{\mu }}\, \gg \, m_{\nu _{e}}\: and\: f_{\nu _{3}}\, \geq \, f_{\nu _{2}}\, \geq \, f_{\nu _{1}}\, . \)Moreover
we choose degenerate Majorana masses for right-handed neutrinos,that is: \( \, M^{ij}_{\nu }\, =\, M_{R}\, \delta ^{ij}\: . \)Finally,without
complex phase in U,the left-handed neutrino mass matrix is:

\begin{equation}
\label{mass}
m_{\nu }\: =\: \frac{1}{M_{R}}\, V^{T}\, \left( \begin{array}{ccc}
f^{2}_{\nu _{1}} & 0 & 0\\
0 & f^{2}_{\nu _{2}} & 0\\
0 & 0 & f^{2}_{\nu _{3}}
\end{array}\right) \, V\, \frac{v^{2}\, \sin ^{2}(\beta )}{2}\; 
\end{equation}
 \( with\; \; V\, =\, \left( \begin{array}{cc}
\cos (\theta ) & \sin (\theta )\\
-\sin (\theta ) & \cos (\theta )
\end{array}\right)  \) for the tau and mu neutrino masses.

Thus,for \( V\neq 1 \) the mass eigenvalues are non-degenerate and very small
for \( \, M_{R}=10^{12}-10^{15}\, Gev\, . \)Consequently,large mixing angles
in V lead,radiatively,to LFV effects in the left-handed slepton mass matrix(even
if the SUSY breaking masses for sleptons are flavor independent at tree level).Indeed,in
minimal SUGRA,the SUSY breaking masses for scalar supersymmetric particles are
universal \( (m_{0}) \) and SUSY breaking parameter are proportional to Yukawa
coupling constants (A ,B ) or masses at the gravitational scale \( (M_{G}) \)
such as:

\[
B^{ij}_{\nu }=M_{\nu _{ij}}\, b0\, ,\, A^{ij}_{\nu }=f_{\nu _{ij}}\, a_{0}\, ,\, A^{ij}_{e}=f_{e_{ij}}\, a_{0}(\; a_{0}\; is\; a\; constant\; ).\]

Then at low energy,radiative corrections to the SUSY breaking parameters are
LFV when Yukawa couplings are LFV.At low energy the LFV off-diagonal components,in
the left-handed slepton mass matrix,are determined by solving the one-loop renormalization
group equations (RGE's) (with boundary conditions so that gaugino masses satisfy
the so-called grand unified theory (GUT) relation at low energy ).A simple approximation
of the LFV off-diagonal components is:

\begin{equation}
\label{off-diag1}
\begin{array}{c}
\left( m^{2}_{\widetilde{L}}\right) _{ij}\: \simeq \, -\frac{1}{8\pi ^{2}}\, \left( 3\, m^{2}_{0}\, +\, a^{2}_{0}\right) \, V^{*}_{ki}\, V_{kj}\, f^{2}_{\nu _{k}}\, \log (\frac{M_{G}}{M_{\nu _{k}}})\\
A^{ij}_{e}\, \simeq \, -\frac{3}{8\pi ^{2}}\, a_{0}\, f_{e_{i}}\, V^{*}_{ki}\, V_{kj}\, f^{2}_{\nu _{k}}\, \log (\frac{M_{G}}{M_{\nu _{k}}})\; \\
\left( m^{2}_{\widetilde{e}}\right) _{ij}\, \simeq \, 0\; for\: i\, \neq \, j\, 
\end{array}
\end{equation}

with \( \; (m^{2}_{\widetilde{L}})\: ,\: A_{e}\: and\: (m^{2}_{\widetilde{e}})\:  \)for
left-left,left-right and right-right components respectively.The right-handed
leptons have one kind of Yukawa interaction,then we have 

\( (m^{2}_{\widetilde{e}})_{ij}\, =0\; for\: i\neq j\: (\, \widetilde{e}\: \; stands\: \; for\: \; right-handed\: \; \; charged\: \; \; slepton\, )\, .\,  \)In
this model we calculate the neutrino radiative decay:\( \, \overline{\nu _{\tau }}\: \longrightarrow \: \nu _{\mu }+\gamma \,  \)and
deduce the transition magnetic moment for different values on SUSY parameters
and for two different Majorana masses.In the next section,we take into account
the \( \, SU(5)RN\,  \) model where\( \, (m^{2}_{\widetilde{e}})_{ij}\, \neq 0\;  \)for
\( i\neq j \).

\subsection{LFV in \protect\( \: SU(5)RN\: \protect \) model}

We consider the supersymmetric \( \, SU(5)\,  \)grand unified theory (GUT)
with right-handed neutrino \cite{6}-\cite{3},as an extension of the MSSMRN,in
which small neutrino mass and large mixing angle are possible without fine-tuning.In
this model,the LFV off-diagonal components \( \, (m^{2}_{\widetilde{e}})_{ij}\,  \)
are different of zero contrary to the MSSMRN case.Then,there is an additional
contribution to the neutrino magnetic moment.The \( \, SU(5)RN\,  \)model has
three families of matter multiplets that is:

\( \Psi _{i}\, the\: quark\: doublet\, , \)\( \Phi _{i}\: the\: charged\: lepton\: singlet\: and\:  \)\( \eta _{i}\: the\: right-handed\: neutrino \)
in 10 ,\( 5^{*} \),1 dimension representations of SU(5) respectively.The higgs
sector contains two \( H\, ,\, \overline{H\, } \)higgs multiplets,in \( \, 5\, and\, 5^{*} \)dimension,such
as \( H=(H_{2},H_{c})\:  \)and \( \overline{H}=(H_{1},\overline{H}_{c}) \)
where \( H_{1}\, ,H_{2}\,  \)are the higgs multiplets of the MSSM and \( \, H_{c}\, ,\, \overline{H}_{c}\, are \)
two different colored higgs multiplets.At the GUT scale \( \, (\, M_{GUT}\approx 3\times 10^{16}\, GEV\, ) \),the
GUT gauge symmetry is spontaneously broken into the SM one: \( SU(5)_{GUT}\, \longrightarrow \, SU(3)_{c}\, \times \, SU(2)_{L}\, \times \, U(1)_{Y}\,  \).So
the superpotential,in the matter sector,above the GUT scale is given as:

\begin{equation}
\label{superpot2}
W\, =\, \frac{1}{4}\, f_{u_{ij}}\, \Psi ^{AB}_{i}\, \Psi ^{CD}_{j}\, H^{E}\, \varepsilon _{ABCDE}\: +\: \sqrt{2}\, f_{d_{ij}}\, \Psi ^{AB}_{i}\, \Phi _{jA}\, \overline{H_{B}}+\: f_{\nu _{ij}}\, \eta _{i}\, \Phi _{jA}\, H^{A}\: +\: \frac{1}{2}\, M_{\eta _{i}\eta _{j}}\, \eta _{i}\, \eta _{j}
\end{equation}
 where A,B,C,D,E are SU(5) indices and run from 1 to 5.At the gravitational
scale (in the minimal SUGRA scenario) the SUSY breaking masses are universal
\( (\, m_{0}\, ) \) and the SUSY breaking parameters are proportional to the
Yukawa coupling constants,as:

\[
A_{u_{ij}}=f_{u_{ij}}\, a_{0}\: ,\: \: A_{d_{ij}}=f_{d_{ij}}\, a_{0}\: \: ,\: \: A_{\nu _{ij}}=f_{\nu _{ij}}\, a_{0\, },\]

with u,d and \( \nu  \) stand for up,down and neutrino respectively.

At the GUT scale,the Yukawa coupling matrices are chosen such as:

\begin{equation}
\label{eq1}
f_{u_{ij}}\, =\, f_{u_{i}}\, \exp (i\, \Phi _{u_{i}})\, \delta _{ij}
\end{equation}

\begin{equation}
\label{eq2}
f_{d_{ij}}\, =\, (\, V^{*}_{KM}\, )_{ik}\, f_{d_{k}}\, (\, V^{T})_{kj}
\end{equation}

\begin{equation}
\label{eq3}
f_{\nu _{ij}}\, =\, f_{\nu _{i}}\, \exp (i\, \Phi _{\nu _{i}})\, \delta _{ij}
\end{equation}

\begin{equation}
\label{eq4}
f_{d_{i}}\, =\, f_{e_{i}}
\end{equation}

where \( V_{KM}\,  \) is the Kobayashi-Maskawa matrix,\( e \) stands for charged
lepton and \( \Phi _{\Psi _{i}}\: (\Psi _{i}=u\, ,\nu \, )\:  \)are phase factors.The
equation (\ref{eq4}) is consistent with the third generation at low energy.For
the two others generations it is different and for simplicity we don't take
them into account.Moreover,at low energy it is possible to write the GUT multiplets
in terms of MSSM fields.Then,we have:

\[
\Psi _{i}=\left\{ \, Q_{i}\, ,\, \exp (-i\, \Phi _{u_{i}}\, \overline{U_{i}}\, ,\, (V_{KM})_{ij}\, \overline{E_{j}}\right\} \; \; ,\; \; \Phi _{i}=\left\{ V_{ij}\, \overline{D_{i}}\, ,\, V_{ij}\, L_{j}\right\} \; \; ,\; \; \eta _{i}=\left\{ \exp (-i\, \Phi _{\nu _{i}})\, \overline{N_{i}}\right\} \]
So the superpotential( \ref{superpot2})becomes:

\begin{equation}
\label{superpot3}
\begin{array}{c}
W\; =\; f_{u_{i}}\, Q_{i}\, \overline{U_{i}}\, H_{2}\: +\: (V^{*}_{KM})_{ij}\, f_{di}\, Q_{i}\, \overline{D_{i}}H_{1}+f_{d_{i}}\overline{E}_{i}L_{i}H_{1}\\
-f_{\nu _{i}}V_{ij}\overline{N}_{i}L_{j}H_{2}+f_{u_{j}}\, (V_{KM})_{ji}\, \overline{E_{i}}\, \overline{U}_{j}H_{c}-\frac{1}{2}f_{u_{i}}\exp (i\Phi _{u_{i}})Q_{i}Q_{i}H_{c}\\
+(V^{*}_{KM})_{ij}f_{d_{j}}\exp (-i\, \Phi _{u_{i}})\overline{U_{i}}\, \overline{D_{j}}\, \overline{H_{c}}-(V^{*}_{KM})_{ij}\, f_{d_{j}}\, Q_{i}\, L_{j}\, \overline{H_{c}}+\: f_{\nu _{i}}\, V_{ij}\, \overline{N_{i}}\, \overline{D_{j}}\, H_{c}\\
+\, \frac{1}{2}\, M_{\nu _{i}\nu _{j}}\, \overline{N_{i}}\, \overline{N_{j}}
\end{array}
\end{equation}

The \( \, f_{u_{i}}\, (V_{KM})_{ij}\, \overline{E_{i}}\, \overline{U_{j}}\, H_{c}\,  \)
in the superpotential (\ref{superpot3}) is not generation diagonal (because
in the GUT,theory leptons and quarks are embedded into the same multiplet).This
mixing gives rise to right-right off-diagonal components \( (\, m^{2}_{\widetilde{e}}\, ) \).The
left-handed neutrino mass matrix is the same as in the MSSMRN case.With \( \, m_{\nu _{\tau }}\, \gg \, m_{\nu _{\mu }}\, \gg \, m_{\nu _{e}}\,  \)
and \( \, f_{\nu _{3}}\, \gg \, f_{\nu _{2}}\, \gg \, f_{\nu _{1}}\,  \) assumptions,the
off-diagonal elements are:

\begin{equation}
\label{off-diag2}
\begin{array}{c}
(\, m^{2}_{\widetilde{e}}\, )_{ij}\, \simeq \, -\frac{3}{8\pi ^{2}}\, f^{2}_{u_{3}}\, (V_{KM})_{3i}\, (V^{*}_{KM})_{3j}\, (\, 3m^{2}_{0}\, +\, a^{2}_{0}\, )\, \log (\frac{M_{G}}{M_{GUT}})\\
(\, m^{2}_{\widetilde{L}}\, )_{ij}\, \simeq \, -\frac{1}{8\pi ^{2}}\, (\, 3m^{2}_{0}\, +\, a^{2}_{0}\, )\, \{\, f^{2}_{\nu _{3}}\, V^{*}_{3i}\, V_{3j}\, \log (\frac{M_{G}}{M_{\nu _{3}}})+\, f^{2}_{\nu _{2}}\, V^{*}_{2i}\, V_{2j}\, \log (\frac{M_{G}}{M_{\nu _{2}}})\, \}\\
A^{ij}_{e}\, \simeq \, -\frac{3}{8\pi ^{2}}\, a_{0}\, \{\, f_{e_{i}}\, V^{*}_{3i}\, V_{3j}\, f^{2}_{\nu _{3}}\, \log (\frac{M_{G}}{M_{\nu _{3}}})+\, f_{e_{i}}\, V^{*}_{2i}\, V_{2j}\, f^{2}_{\nu _{2}}\, \log (\frac{M_{G}}{M_{\nu _{2}}})\\
+\, 3\, f_{e_{j}}\, (V^{*}_{KM)_{3j}}\, (V_{KM})_{3i}\, f^{2}_{u_{3}}\, \log (\frac{M_{G}}{M_{GUT}})\, \}
\end{array}
\end{equation}

So in the \( SU(5)RN \) case,we have the \( (m^{2}_{\widetilde{e}})_{ij} \)
contribution to the transition magnetic moment.Also,due to this contribution,we
can hope an increase of the transition magnetic moment value.We discuss about
it in the next section.

\section{Amplitudes contributing to the decay : \protect\( \overline{\nu _{\tau }}\: \longrightarrow \: \nu _{\mu }+\gamma \, \protect \)}

\subsection{Amplitudes contributing in the MSSMRN model}

The radiative decay of neutrino consists of the process:\( \nu _{3}\, \longrightarrow \, \nu _{2}\, +\, \gamma  \)
where \( \nu _{3}\: and\: \nu _{2}\:  \)are mass eigenstates (essentially left-handed)
very light,so that \( m_{\nu }\approx 1ev\: ,\: \nu _{3}\:  \) being the heaviest
one.To fix ideas,we take \( \nu _{3} \) as predominantly \( \nu _{\tau } \)
and \( \nu _{2} \) predominantly \( \nu _{\mu } \).The MSSMRN Feynman rules,used
for the calculations of the amplitudes,can be found in reference \cite{5} and
in Annex A.In Annex B,we have the Feynman integrals.The tau mass is greater
than muon one,the same for neutrino masses,Then we have only considered the
decay:\( \overline{\nu _{\tau }}\: \longrightarrow \: \nu _{\mu }+\gamma \,  \)
which is dominant compare to \( \nu _{\tau }\: \longrightarrow \: \overline{\nu _{\mu }}+\gamma \,  \).There
are two kinds of diagrams contributing to the decay :

+ the left-left non-diagonal mass insertion diagrams illustrated in figures(\ref{f1}).However,on
the diagrams \ref{f1}(b) and \ref{f1}(c)we have a flavor conserving left-right
mass insertion.

+ and the left-right non diagonal mass insertion diagrams represented in figures(\ref{f2})

\subsubsection{Left-left non-diagonal mass insertions}

\begin{figure}
{\par\centering \begin{picture}(195,120)(0,0)\par}

{\par\centering \ArrowLine(0,20)(40,20)\Text(20,15)[tc]{\(\overline{\nu}_{\tau}\)}\Text(20,25)[bc]{\(p\)}\par}

{\par\centering \Vertex(40,20){2}\par}

{\par\centering \Text(38,15)[tc]{\({\nu}_{\tau}\)}\par}

{\par\centering \DashLine(95,20)(40,20){3}\Text(57,15)[tc]{\(\tilde{\tau}_{L}\)}\par}

{\par\centering \Vertex(80,20){2}\par}

{\par\centering \Text(80,0)[tc]{\ref{f1}(a1)}\Text(80,24)[bc]{\(\overline{m}^{2}_{32}\)}\par}

{\par\centering \DashArrowLine(120,20)(80,20){3}\Text(100,15)[tc]{\(\tilde {\mu}_{L}\)}\par}

{\par\centering \Vertex(120,20){2}\par}

{\par\centering \ArrowLine(120,20)(160,20)\Text(140,15)[tc]{\({\nu}_{\mu}\)}\Text(140,25)[bc]{\(p-q\)}\par}

{\par\centering \CArc(80,20)(40,0,180)\par}

{\par\centering \Line(27,17)(33,23)\Line(27,23)(33,17)\Text(80,65)[bc]{\(\tilde{W}^{+}\)}\par}

{\par\centering \Photon(115,40)(120,60){-3}{4}\Text(125,0)[bl]{\(q\)}\par}

{\par\centering \end{picture}\hskip2cm \begin{picture}(160,80)(0,0)
\ArrowLine(0,20)(40,20)\Text(20,15)[tc]{\(\overline{\nu}_{\tau}\)}\Text(20,25)[bc]{\(p\)}
\Vertex(40,20){2}\par}

{\par\centering \Text(38,15)[tc]{\({\nu}_{\tau}\)}
\DashLine(85,20)(40,20){3}\Text(60,15)[tc]{\(\tilde {\tau}_{L}\)}
\Vertex(80,20){2}
\Text(80,0)[tc]{\ref{f1}(a2)}\Text(80,24)[bc]{\(\overline{m}^{2}_{32}\)}
\DashArrowLine(120,20)(80,20){3}\Text(100,15)[tc]{\(\tilde {\mu}_{L}\)}
\Vertex(120,20){2}
\ArrowLine(120,20)(160,20)\Text(140,15)[tc]{\({\nu}_{\mu}\)}\Text(140,25)[bc]{\(p-q\)}
\CArc(80,20)(40,0,180)
\Line(27,17)(33,23)\Line(27,23)(33,17) 
\Text(80,65)[bc]{\(\tilde{W}^{+}\)}
\Photon(110,20)(120,0){3}{4}\Text(125,0)[bl]{\(q\)}
\end{picture}\par}

{\par\centering \begin{picture}(190,110)(0,0) \ArrowLine(0,20)(40,20)\Text(20,15)[tc]{\(\overline{\nu}_{\tau}\)}\Text(26,33)[b c]{\(\tilde{H}^{+}_{1L}\)}\par}

{\par\centering \Vertex(40,20){2}\par}

{\par\centering \DashLine(95,20)(40,20){3} \Text(50,15)[tc]{\(\tilde {\tau}_{R}\)} \Vertex(57,20){2}\par}

{\par\centering \Text(75,15)[tc]{\(\tilde {\tau}_{L}\)}\par}

{\par\centering \Vertex(88,20){2}\par}

{\par\centering \Text(91,24)[bc]{\(\overline{m}^{2}_{32}\)}\par}

{\par\centering \Text(80,0)[tc]{\ref{f1}(b1)}\Text(57,24)[bc]{\(\overline{m}^{2}_{LR{\tau}}\)}\par}

{\par\centering \DashArrowLine(120,20)(80,20){3}\Text(110,15)[tc]{\(\tilde{\mu}_{L}\)}\par}

{\par\centering \Vertex(120,20){2}\par}

{\par\centering \ArrowLine(120,20)(160,20)\Text(140,15)[tc]{\({\nu}_{\mu}\)}\Text(138,38)[bc] {\(\tilde{W}^{+}_{L}\)}\par}

{\par\centering \CArc(80,20)(40,0,60)\par}

{\par\centering \Text(75,71)[bc]{\(\tilde{W}^{+}_{R}\)}\par}

{\par\centering \Vertex(100,55){2}\par}

{\par\centering \Text(100,60)[bc]{\({M}_{2}\)}\par}

{\par\centering \CArc(80,20)(40,60,180)\par}

{\par\centering \Vertex(55,52){2}\par}

{\par\centering \Text(45,55)[bc]{\(M_{cos}\)}\par}

{\par\centering \Photon(115,40)(120,60){-3}{4}\Text(125,0)[bl]{\(q\)}\par}

{\par\centering \end{picture}\hskip2cm \begin{picture}(170,80)(0,0) \ArrowLine(0,20)(40,20)\Text(20,15)[tc]{\(\overline{\nu}_{\tau}\)}\Text(26,33)[bc]{\(\tilde{H}^{+}_{1L}\)}\par}

{\par\centering \Vertex(40,20){2}\par}

{\par\centering \DashLine(95,20)(40,20){3} \Text(50,15)[tc]{\(\tilde {\tau}_{R}\)} \Vertex(57,20){2}\par}

{\par\centering \Text(75,15)[tc]{\(\tilde {\tau}_{L}\)}\par}

{\par\centering \Vertex(88,20){2}\par}

{\par\centering \Text(91,24)[bc]{\(\overline{m}^{2}_{32}\)}\par}

{\par\centering \Text(80,0)[tc]{\ref{f1}(b2)}\Text(57,24)[bc]{\(\overline{m}^{2}_{LR{\tau}}\)}\par}

{\par\centering \DashArrowLine(120,20)(80,20){3}\Text(100,15)[tc]{\(\tilde{\mu}_{L}\)}\par}

{\par\centering \Vertex(120,20){2}\par}

{\par\centering \ArrowLine(120,20)(160,20)\Text(140,15)[tc]{\({\nu}_{\mu}\)}\Text(138,38)[bc]{\(\tilde{W}^{+}_{L}\)}\par}

{\par\centering \CArc(80,20)(40,0,60)\par}

{\par\centering \Text(75,71)[bc]{\(\tilde{W}^{+}_{R}\)}\par}

{\par\centering \Vertex(100,55){2}\par}

{\par\centering \Text(100,60)[bc]{\({M}_{2}\)}\par}

{\par\centering \CArc(80,20)(40,60,180)\par}

{\par\centering \Vertex(55,52){2}\par}

{\par\centering \Text(45,55)[bc]{\(M_{cos}\)}\par}

{\par\centering \Photon(110,20)(120,0){-3}{4}\Text(125,0)[bl]{\(q\)}\par}

{\par\centering \end{picture}\par}

{\par\centering \begin{picture}(190,110)(0,0) \ArrowLine(0,20)(40,20)\Text(20,15)[tc]{\(\overline{\nu}_{\tau}\)}\Text(26,33)[bc]{\(\tilde{H}^{+}_{1L}\)}\par}

{\par\centering \Vertex(40,20){2}\par}

{\par\centering \DashLine(95,20)(40,20){3} \Text(50,15)[tc]{\(\tilde {\tau}_{R}\)} \Vertex(57,20){2}\par}

{\par\centering \Text(75,15)[tc]{\(\tilde {\tau}_{L}\)}\par}

{\par\centering \Vertex(88,20){2}\par}

{\par\centering \Text(91,24)[bc]{\(\overline{m}^{2}_{32}\)}\par}

{\par\centering \Text(80,0)[tc]{\ref{f1}(c1)}\Text(57,24)[bc]{\(\overline{m}^{2}_{LR{\tau}}\)}\par}

{\par\centering \DashArrowLine(120,20)(80,20){3}\Text(110,15)[tc]{\(\tilde{\mu}_{L}\)}\par}

{\par\centering \Vertex(120,20){2}\par}

{\par\centering \ArrowLine(120,20)(160,20)\Text(140,15)[tc]{\({\nu}_{\mu}\)}\Text(138,38)[bc]{\(\tilde{W}^{+}_{L}\)}\par}

{\par\centering \CArc(80,20)(40,0,60)\par}

{\par\centering \Text(75,71)[bc]{\(\tilde{H}^{+}_{2R}\)}\par}

{\par\centering \Vertex(100,55){2}\par}

{\par\centering \Text(100,60)[bc]{\({M}_{sin}\)}\par}

{\par\centering \CArc(80,20)(40,60,180)\par}

{\par\centering \Vertex(55,52){2}\par}

{\par\centering \Text(45,55)[bc]{\({\mu}\)}\par}

{\par\centering \Photon(115,40)(120,60){-3}{4}\Text(125,0)[bl]{\(q\)}\par}

{\par\centering \end{picture}\hskip2cm \begin{picture}(170,80)(0,0) \ArrowLine(0,20)(40,20)\Text(20,15)[tc]{\(\overline{\nu}_{\tau}\)}\Text(26,33)[bc]{\(\tilde{H}^{+}_{1L}\)}\par}

{\par\centering \Vertex(40,20){2}\par}

{\par\centering \DashLine(95,20)(40,20){3} \Text(50,15)[tc]{\(\tilde {\tau}_{R}\)} \Vertex(57,20){2}\par}

{\par\centering \Text(75,15)[tc]{\(\tilde {\tau}_{L}\)}\par}

{\par\centering \Vertex(88,20){2}\par}

{\par\centering \Text(91,24)[bc]{\(\overline{m}^{2}_{32}\)}\par}

{\par\centering \Text(80,0)[tc]{\ref{f1}(c2)}\Text(57,24)[bc]{\(\overline{m}^{2}_{LR{\tau}}\)}\par}

{\par\centering \DashArrowLine(120,20)(80,20){3}\Text(100,15)[tc]{\(\tilde{\mu}_{L}\)}\par}

{\par\centering \Vertex(120,20){2}\par}

{\par\centering \ArrowLine(120,20)(160,20)\Text(140,15)[tc]{\({\nu}_{\mu}\)}\par}

{\par\centering \Text(138,38)[bc]{\(\tilde{W}^{+}_{L}\)}\par}

{\par\centering \CArc(80,20)(40,0,60)\par}

{\par\centering \Text(75,71)[bc]{\(\tilde{H}^{+}_{2R}\)}\par}

{\par\centering \Vertex(100,55){2}\par}

{\par\centering \Text(100,60)[bc]{\({M}_{sin}\)}\par}

{\par\centering \CArc(80,20)(40,60,180)\par}

{\par\centering \Vertex(55,52){2}\par}

{\par\centering \Text(45,55)[bc]{\({\mu}\)}\par}

{\par\centering \Photon(110,20)(120,0){-3}{4}\Text(125,0)[bl]{\(q\)}\par}

{\par\centering \end{picture}\par}

\caption{\label{f1}Chargino contributions with photon \protect\( (\gamma )\protect \)
and charged sleptons \protect\( (\, \widetilde{\tau },\widetilde{\mu }\, )\protect \)
coupling to left-left non-diagonal mass insertion (\protect\( (\overline{m}^{2})_{32}\protect \)
),where \protect\( M_{cos}\, =\, \sqrt{2}\, m_{w}\, \cos (\beta )\protect \)
and \protect\( M_{sin}\, =\, \sqrt{2}\, m_{w}\, \sin (\beta )\protect \).In
(b) and (c) there is a flavor diagonal left-right insertion \protect\( \overline{m}^{2}_{LR\tau }\protect \)(L
stands for left and R for right)with \protect\( \widetilde{W}\protect \) and
\protect\( \widetilde{H}\protect \) for wino,higgsino components of chargino.\protect\( \mu \, \protect \)
and \protect\( M_{2\, }\protect \) are higgsino and wino masses respectively.}
\end{figure}
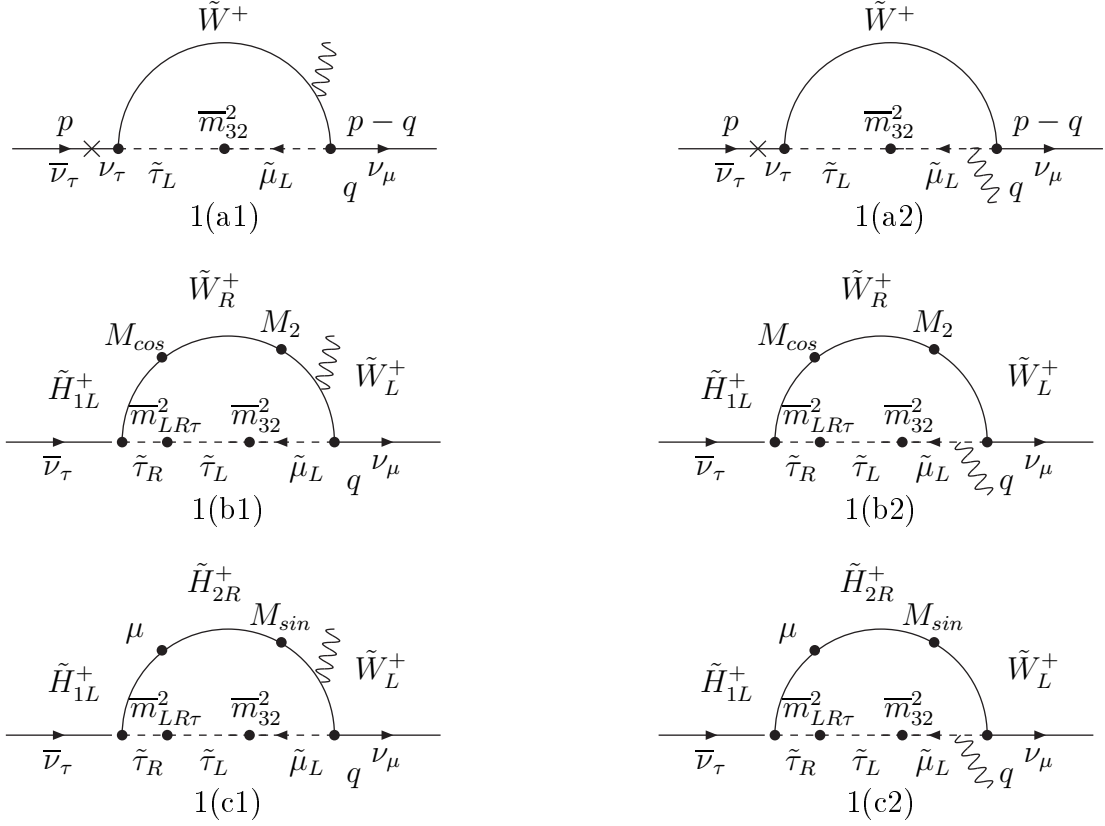

Due to the photon-chargino coupling,there are 6 diagrams illustrated in figures
(\ref{f1}).The diagrams (1a1) and (1a2) have no left-right diagonal insertion

From the diagrams \ref{f1}(a) we have:

\( T_{1_{a}}\: =\: \frac{i\, g^{2}\, e}{16\, \pi ^{2}}\, \frac{(\overline{m}^{2})_{32}}{[\overline{m}_{\widetilde{\tau _{L}}}^{2}\, -\overline{m}_{\widetilde{\mu _{L}}}^{2}\, ]}\: \overline{U}_{\nu _{\mu }}\, (p-q)\, P_{R}\, \sigma ^{\mu \nu }\, \varepsilon _{\mu (q)}\, q_{\nu }\, \times \left[ F_{h}\, (\, X_{\widetilde{\tau }_{L}}\, ,\, X_{\widetilde{\mu }_{L}})\, P\: +\: G_{h}\, (\, X_{\widetilde{\tau }_{L}}\, ,\, X_{\widetilde{\mu }_{L}}\, )\, q\right] \, U_{\nu _{\tau }}(p) \)

with \( F_{h}\, (\, X_{\widetilde{\tau }_{L}}\, ,\, X_{\widetilde{\mu }_{L}}\, )\, =\, \frac{1}{\overline{m}_{\widetilde{\tau }_{L}}^{2}}\, \left[ f(X_{\widetilde{\tau }_{L}})\, +h(X_{\widetilde{\tau }_{L}})\right] \: -\: \frac{1}{\overline{m}_{\widetilde{\mu }_{L}}^{2}}\, \left[ f(X_{\widetilde{\mu }_{L}})+h(X_{\widetilde{\mu _{L}}})\right] ,\; P_{R}\; right\; chirality\; projector \)

\( and\: G_{h}\, (\, X_{\widetilde{\tau }_{L}}\, ,\, X_{\widetilde{\mu }_{L}}\, )\, =\, \frac{1}{\overline{m}_{\widetilde{\tau }_{L}}^{2}}\, \left[ g(X_{\widetilde{\tau }_{L}})\, +\frac{1}{2}\, h(X_{\widetilde{\tau }_{L}})\right] \: -\: \frac{1}{\overline{m}_{\widetilde{\mu }_{L}}^{2}}\, \left[ g(X_{\widetilde{\mu }_{L}})+\frac{1}{2}\, h(X_{\widetilde{\mu _{L}}})\right]  \)

where \( X_{\widetilde{\tau }_{L}}\, =\, \frac{M^{2}_{2}}{\overline{m}_{\widetilde{\tau }_{L}}^{2}} \)
and \( X_{\widetilde{\mu }_{L}}\, =\, \frac{M^{2}_{2}}{\overline{m}_{\widetilde{\mu }_{L}}^{2}} \)
.\( M_{2} \),\( \overline{m}_{\widetilde{\tau }_{L}}^{2} \) and \( \overline{m}_{\widetilde{\mu }_{L}}^{2} \)
are the wino mass,the diagonal components of the stau and the smuon mass matrix.q
is the outgoing photon momentum and p is the neutrino tau one.the functions
\( f(X)\; ,\; g(X)\; and\; h(X)\:  \) are given in annexB.\( (\overline{m}^{2})_{32} \)
stands for left-left non-diagonal mass insertion for third and second generation.

As,in the previous section,we suppose neutrino mass hierarchy: \( \, m_{\nu _{\tau }}\, \gg \, m_{\nu _{\mu }}\, \gg \, m_{\nu _{e}}\,  \)and
\( \, f_{\nu _{3}}\, \gg \, f_{\nu _{2}}\, \gg \, f_{\nu _{1}}\,  \).From the
atmospheric neutrino observation,we take \( V_{32}\:  \)to be of order of unit.Then,from
the equation(\ref{off-diag1}) we have:

\begin{equation}
\label{In1}
(\overline{m}^{2})_{32}\, \simeq \, -\frac{1}{8\pi ^{2}}\, (3\, m^{2}_{0}\, +\, a^{2}_{0})\, V_{33}^{*}V_{32}\, f^{2}_{\nu _{3}}\, \log (\frac{M_{G}}{M_{\nu _{3}}})
\end{equation}
 Due to the higgsino vertex (proportional to the neutrino mass)the diagrams
\ref{f1}(1b) and (1c) show a left-right diagonal mass insertion \( (\overline{m}_{LR}^{2})_{\tau } \).The
explicit expression is: 
\begin{equation}
\label{In2}
(\overline{m}_{LR}^{2})_{\tau }\, =\, m_{\tau }\, \mu \, \tan (\beta )\, -\, a_{0}\, m_{0}\, m_{\tau }\; \; \; \; ,\; \tau \; for\; lepton\; tau
\end{equation}
 The amplitudes associated to these diagrams are:

\[
\begin{array}{c}
\\
T_{1b}\, =\, \frac{-\, i\, g_{2}^{2}\, e\, m_{\nu _{\tau }}\, M_{2}\, (\overline{m}_{LR}^{2})_{\tau }\, (\overline{m}^{2})_{32}\, }{16\, \pi ^{2}\, [M_{2}^{2}\, -\, \mu ^{2}]\, [\overline{m}^{2}_{\widetilde{\tau _{R}}}\, -\overline{m}_{\widetilde{\tau }_{L}}^{2}]}\, \times \, \sum ^{n=3}_{n=0}\, \frac{(-1)^{n}}{[\overline{m}_{n}^{2}\, -\overline{m}_{\widetilde{\mu }_{L}}^{2}]}\, \frac{1}{\overline{M}_{n}^{2}}\, \times \; \; \; \; \; \; \; \; \; \; \; \; \; \; \; \; \; \; \; \; \; \\
\; \; \; \; \; \; \; \; \; \; \; \; \; \; \; \; \left\{ A_{1b1}(X_{\widetilde{w}_{n}}\, ,\, X_{\widetilde{\mu }_{n}})\, +\, A_{1b2}(X_{\widetilde{w}_{n}}\, ,\, X_{\widetilde{\mu }_{n}})\right\} \, \overline{U}_{\nu _{\mu }}(p-q)\, P_{R}\, \sigma ^{\mu \nu }\, \varepsilon _{\mu }(q)\, q^{\nu }\, U_{\nu _{\tau }}(p)\\

\end{array}\]

\[
\begin{array}{c}
\\
T_{1c}\, =\, \frac{-\, i\, g_{2}^{2}\, e\, m_{\nu _{\tau }}\, \mu \: \tan (\beta )\, (\overline{m}_{LR}^{2})_{\tau }\, (\overline{m}^{2})_{32}\, }{16\, \pi ^{2}\, [M_{2}^{2}\, -\, \mu ^{2}]\, [\overline{m}^{2}_{\widetilde{\tau _{R}}}\, -\overline{m}_{\widetilde{\tau }_{L}}^{2}]}\, \times \, \sum ^{n=3}_{n=0}\, \frac{(-1)^{n}}{[\overline{m}_{n}^{2}\, -\overline{m}_{\widetilde{\mu }_{L}}^{2}]}\, \frac{1}{\overline{M}_{n}^{2}}\, \times \; \; \; \; \; \; \; \; \; \; \; \; \; \; \; \; \; \; \; \; \; \\
\; \; \; \; \; \; \; \; \; \; \; \; \; \; \; \; \left\{ A_{1c1}(X_{\widetilde{w}_{n}}\, ,\, X_{\widetilde{\mu }_{n}})\, +\, A_{1c2}(X_{\widetilde{w}_{n}}\, ,\, X_{\widetilde{\mu }_{n}})\right\} \, \overline{U}_{\nu _{\mu }}(p-q)\, P_{R}\, \sigma ^{\mu \nu }\, \varepsilon _{\mu }(q)\, q^{\nu }\, U_{\nu _{\tau }}(p)\\

\end{array}\]

\[
\begin{array}{c}
\\
\overline{m}_{n}^{2}\: =\, \left\{ \begin{array}{c}
\overline{m}_{\widetilde{\tau }_{R}}^{2}\; ,\; for\; n\, =0,1\\
\overline{m}_{\widetilde{\tau }_{L}}^{2}\; ,\; for\; n\, =2,3
\end{array}\right. \: \: \: \; \; ,\; \; \; \; \overline{M}_{n}^{2}\: =\, \left\{ \begin{array}{c}
\overline{m}_{\widetilde{\tau }_{R}}^{2}\; ,\; for\; n\, =0\\
\overline{m}_{\widetilde{\tau }_{L}}^{2}\; ,\; for\; n\, =3\\
\overline{m}_{\widetilde{\mu }_{L}}^{2}\; ,\; for\; n\, =1,2
\end{array}\right. \\
\\
and\; \; \; \; X_{\widetilde{\mu }_{n}}\: =\, \frac{\mu ^{2}}{\overline{M}_{n}^{2}}\; \; \; \; \; \; \; \; \; \; \; \; ,\; \; \; \; \; \; \; \; \; \; X_{\widetilde{w}_{n}}\: =\, \frac{M_{2}^{2}}{\overline{M}_{n}^{2}}\: \: \: \: \: \: (\: \mu \: =\: higgsino\: \: \: \: \: mass\: )\: .
\end{array}\]

The explicit expressions of the functions \( \: A_{1b1}\: ,\: A_{1b2}\: ,\: A_{1c1}\: ,\: A_{1c2} \)
are given in annexB.

\subsubsection{Left-right non-diagonal mass insertion}

As left-left case we have six diagrams represented in figures( \ref{f2}).The
left-right non-diagonal mass insertion (\ref{off-diag1}) is:

\begin{equation}
\label{In3}
A_{e_{32}}\, \simeq \, -\frac{3}{8\pi ^{2}}\, a_{0}\, f_{e_{3}}\, V_{33}^{*}V_{32}\, f^{2}_{\nu _{3}}\, \log (\frac{M_{G}}{M_{\nu _{3}}}).
\end{equation}

Then the amplitudes are:

+ for the diagrams (2a):

\( \begin{array}{c}
\\
T_{2a}\, =\, \frac{i\, g_{2}^{2}\, e\, (\overline{m}_{LR}^{2})_{\tau }\, A_{e_{32}}\, }{16\, \pi ^{2}\, [\overline{m}^{2}_{\widetilde{\tau _{R}}}\, -\overline{m}_{\widetilde{\tau }_{L}}^{2}]}\, \times \, \sum ^{n=3}_{n=0}\, \frac{(-1)^{n}}{[\overline{m}_{n}^{2}\, -\overline{m}_{\widetilde{\mu }_{L}}^{2}]}\, \frac{1}{\overline{M}_{n}^{2}}\, \times \; \; \; \; \; \; \; \; \; \; \; \; \; \; \; \; \; \; \; \; \; \\
\; \, \overline{U}_{\nu _{\mu }}(p-q)\, P_{R}\, \sigma ^{\mu \nu }\, \varepsilon _{\mu }(q)\, q^{\nu }\, \left\{ P\, [h(X_{\widetilde{w}_{n}})\, +f(X_{\widetilde{w}_{n}})]\, +\, q\, [h(X_{\widetilde{w}_{n}})\, +\, g(X_{\widetilde{w}_{n}})]\right\} \, U_{\nu _{\tau }}(p)\\

\end{array} \) 

For the functions h and g see annex B for \( f_{e_{3}} \) annex A.

+ for diagrams (2b):

\begin{itemize}
\item 
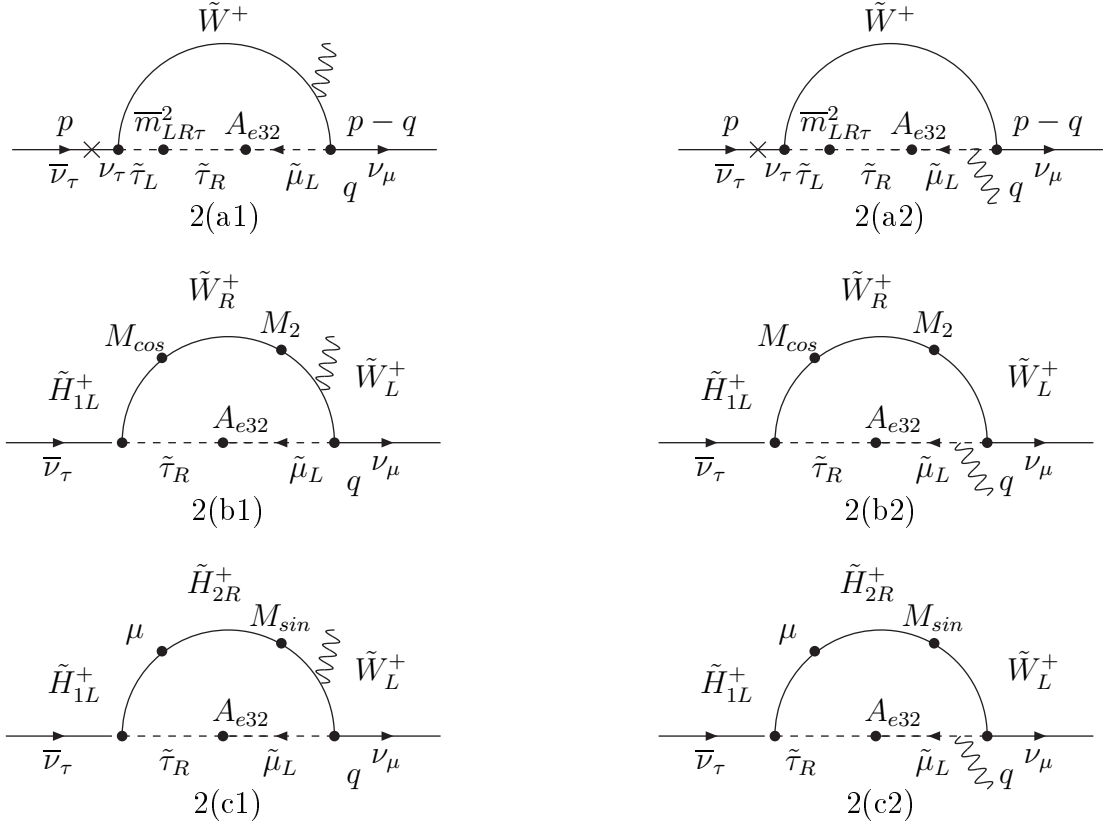
\begin{figure}
{\par\centering \begin{picture}(195,120)(0,0)\par}

{\par\centering \ArrowLine(0,20)(40,20)\Text(20,15)[tc]{\(\overline{\nu}_{\tau}\)}\Text(20,25)[bc]{\(p\)}\par}

{\par\centering \Vertex(40,20){2}\par}

{\par\centering \Text(38,15)[tc]{\({\nu}_{\tau}\)}\par}

{\par\centering \DashLine(95,20)(40,20){3}\Text(50,15)[tc]{\(\tilde{\tau}_{L}\)}\par}

{\par\centering \Vertex(57,20){2}\par}

{\par\centering \Text(75,15)[tc]{\(\tilde{\tau}_{R}\)}\par}

{\par\centering \Vertex(88,20){2}\par}

{\par\centering \Text(91,24)[bc]{\({A}_{e32}\)}\par}

{\par\centering \Text(80,0)[tc]{\ref{f2}(a1)}\Text(60,24)[bc]{\(\overline{m}^{2}_{LR{\tau}}\)}\par}

{\par\centering \DashArrowLine(120,20)(80,20){3}\Text(110,15)[tc]{\(\tilde{\mu}_{L}\)}\par}

{\par\centering \Vertex(120,20){2}\par}

{\par\centering \ArrowLine(120,20)(160,20)\Text(140,15)[tc]{\({\nu}_{\mu}\)}\par}

{\par\centering \Text(140,25)[bc]{\(p-q\)}\par}

{\par\centering \CArc(80,20)(40,0,180)\par}

{\par\centering \Line(27,17)(33,23)\Line(27,23)(33,17)\par}

{\par\centering \Text(80,65)[bc]{\(\tilde{W}^{+}\)}\par}

{\par\centering \Photon(115,40)(120,60){-3}{4}\Text(125,0)[bl]{\(q\)}\par}

{\par\centering \end{picture}\hskip2cm \begin{picture}(160,80)(0,0)
\ArrowLine(0,20)(40,20)\Text(20,15)[tc]{\(\overline{\nu}_{\tau}\)}\Text(20,25)[bc]{\(p\)}
\Vertex(40,20){2}\par}

{\par\centering \Text(38,15)[tc]{\({\nu}_{\tau}\)}
\DashLine(95,20)(40,20){3}\Text(50,15)[tc]{\(\tilde {\tau}_{L}\)}
\Vertex(57,20){2}\par}

{\par\centering \Text(75,15)[tc]{\(\tilde {\tau}_{R}\)}\par}

{\par\centering \Vertex(88,20){2}\par}

{\par\centering \Text(91,24)[bc]{\({A}_{e32}\)}
\Text(80,0)[tc]{\ref{f2}(a2)}\Text(60,24)[bc]{\(\overline{m}^{2}_{LR{\tau}}\)}
\DashArrowLine(120,20)(80,20){3}\Text(100,15)[tc]{\(\tilde {\mu}_{L}\)}
\Vertex(120,20){2}
\ArrowLine(120,20)(160,20)\Text(140,15)[tc]{\({\nu}_{\mu}\)}\Text(140,25)[bc]{\(p-q\)}
\CArc(80,20)(40,0,180)
\Line(27,17)(33,23)\Line(27,23)(33,17) 
\Text(80,65)[bc]{\(\tilde{W}^{+}\)}
\Photon(110,20)(120,0){3}{4}\Text(125,0)[bl]{\(q\)}
\end{picture}\par}

{\par\centering \begin{picture}(190,110)(0,0) \ArrowLine(0,20)(40,20)\Text(20,15)[tc]{\(\overline{\nu}_{\tau}\)}\Text(26,33)[b c]{\(\tilde{H}^{+}_{1L}\)}\par}

{\par\centering \Vertex(40,20){2}\par}

{\par\centering \DashLine(95,20)(40,20){3} \Text(60,15)[tc]{\(\tilde {\tau}_{R}\)} \Vertex(78,20){2}\par}

{\par\centering \Text(85,24)[bc]{\({A}_{e32}\)}\par}

{\par\centering \Text(80,0)[tc]{\ref{f2}(b1)}\par}

{\par\centering \DashArrowLine(120,20)(80,20){3}\Text(110,15)[tc]{\(\tilde{\mu}_{L}\)}\par}

{\par\centering \Vertex(120,20){2}\par}

{\par\centering \ArrowLine(120,20)(160,20)\Text(140,15)[tc]{\({\nu}_{\mu}\)}\Text(138,38)[bc] {\(\tilde{W}^{+}_{L}\)}\par}

{\par\centering \CArc(80,20)(40,0,60)\par}

{\par\centering \Text(75,71)[bc]{\(\tilde{W}^{+}_{R}\)}\par}

{\par\centering \Vertex(100,55){2}\par}

{\par\centering \Text(100,60)[bc]{\({M}_{2}\)}\par}

{\par\centering \CArc(80,20)(40,60,180)\par}

{\par\centering \Vertex(55,52){2}\par}

{\par\centering \Text(45,55)[bc]{\(M_{cos}\)}\par}

{\par\centering \Photon(115,40)(120,60){-3}{4}\Text(125,0)[bl]{\(q\)}\par}

{\par\centering \end{picture}\hskip2cm \begin{picture}(170,80)(0,0) \ArrowLine(0,20)(40,20)\Text(20,15)[tc]{\(\overline{\nu}_{\tau}\)}\Text(26,33)[bc]{\(\tilde{H}^{+}_{1L}\)}\par}

{\par\centering \Vertex(40,20){2}\par}

{\par\centering \DashLine(95,20)(40,20){3} \Text(60,15)[tc]{\(\tilde {\tau}_{R}\)} \Vertex(78,20){2}\par}

{\par\centering \Text(85,24)[bc]{\({A}_{e32}\)}\par}

{\par\centering \Text(80,0)[tc]{\ref{f2}(b2)}\par}

{\par\centering \DashArrowLine(120,20)(80,20){3}\Text(100,15)[tc]{\(\tilde{\mu}_{L}\)}\par}

{\par\centering \Vertex(120,20){2}\par}

{\par\centering \ArrowLine(120,20)(160,20)\Text(140,15)[tc]{\({\nu}_{\mu}\)}\Text(138,38)[bc]{\(\tilde{W}^{+}_{L}\)}\par}

{\par\centering \CArc(80,20)(40,0,60)\par}

{\par\centering \Text(75,71)[bc]{\(\tilde{W}^{+}_{R}\)}\par}

{\par\centering \Vertex(100,55){2}\par}

{\par\centering \Text(100,60)[bc]{\({M}_{2}\)}\par}

{\par\centering \CArc(80,20)(40,60,180)\par}

{\par\centering \Vertex(55,52){2}\par}

{\par\centering \Text(45,55)[bc]{\(M_{cos}\)}\par}

{\par\centering \Photon(110,20)(120,0){-3}{4}\Text(125,0)[bl]{\(q\)}\par}

{\par\centering \end{picture}\par}

{\par\centering \begin{picture}(190,110)(0,0) \ArrowLine(0,20)(40,20)\Text(20,15)[tc]{\(\overline{\nu}_{\tau}\)}\Text(26,33)[bc]{\(\tilde{H}^{+}_{1L}\)}\par}

{\par\centering \Vertex(40,20){2}\par}

{\par\centering \DashLine(95,20)(40,20){3} \Text(60,15)[tc]{\(\tilde {\tau}_{R}\)} \Vertex(78,20){2}\par}

{\par\centering \Text(85,24)[bc]{\({A}_{e32}\)}\par}

{\par\centering \Text(80,0)[tc]{\ref{f2}(c1)}\par}

{\par\centering \DashArrowLine(120,20)(80,20){3}\Text(100,15)[tc]{\(\tilde{\mu}_{L}\)}\par}

{\par\centering \Vertex(120,20){2}\par}

{\par\centering \ArrowLine(120,20)(160,20)\Text(140,15)[tc]{\({\nu}_{\mu}\)}\Text(138,38)[bc]{\(\tilde{W}^{+}_{L}\)}\par}

{\par\centering \CArc(80,20)(40,0,60)\par}

{\par\centering \Text(75,71)[bc]{\(\tilde{H}^{+}_{2R}\)}\par}

{\par\centering \Vertex(100,55){2}\par}

{\par\centering \Text(100,60)[bc]{\({M}_{sin}\)}\par}

{\par\centering \CArc(80,20)(40,60,180)\par}

{\par\centering \Vertex(55,52){2}\par}

{\par\centering \Text(45,55)[bc]{\({\mu}\)}\par}

{\par\centering \Photon(115,40)(120,60){-3}{4}\Text(125,0)[bl]{\(q\)}\par}

{\par\centering \end{picture}\hskip2cm \begin{picture}(170,80)(0,0) \ArrowLine(0,20)(40,20)\Text(20,15)[tc]{\(\overline{\nu}_{\tau}\)}\Text(26,33)[bc]{\(\tilde{H}^{+}_{1L}\)}\par}

{\par\centering \Vertex(40,20){2}\par}

{\par\centering \DashLine(95,20)(40,20){3} \Text(50,15)[tc]{\(\tilde {\tau}_{R}\)} \Vertex(78,20){2}\par}

{\par\centering \Text(85,24)[bc]{\({A}_{e32}\)}\par}

{\par\centering \Text(80,0)[tc]{\ref{f2}(c2)}\par}

{\par\centering \DashArrowLine(120,20)(80,20){3}\Text(100,15)[tc]{\(\tilde{\mu}_{L}\)}\par}

{\par\centering \Vertex(120,20){2}\par}

{\par\centering \ArrowLine(120,20)(160,20)\Text(140,15)[tc]{\({\nu}_{\mu}\)}\par}

{\par\centering \Text(138,38)[bc]{\(\tilde{W}^{+}_{L}\)}\par}

{\par\centering \CArc(80,20)(40,0,60)\par}

{\par\centering \Text(75,71)[bc]{\(\tilde{H}^{+}_{2R}\)}\par}

{\par\centering \Vertex(100,55){2}\par}

{\par\centering \Text(100,60)[bc]{\({M}_{sin}\)}\par}

{\par\centering \CArc(80,20)(40,60,180)\par}

{\par\centering \Vertex(55,52){2}\par}

{\par\centering \Text(45,55)[bc]{\({\mu}\)}\par}

{\par\centering \Photon(110,20)(120,0){-3}{4}\Text(125,0)[bl]{\(q\)}\par}

{\par\centering \end{picture}\par}

\caption{\label{f2}Chargino contributions with photon \protect\( (\gamma )\protect \)
and charged sleptons \protect\( (\, \widetilde{\tau },\widetilde{\mu }\, )\protect \)
coupling to left-right non-diagonal mass insertion (\protect\( A_{e32}\protect \)).}
\end{figure}

\end{itemize}

\[
\begin{array}{c}
\\
T_{2b}\, =\, \frac{i\, g_{2}^{2}\, e\, m_{\nu _{\tau }}\, M_{2}\, A_{e_{32}}\, }{16\, \pi ^{2}\, [M_{2}^{2}\, -\, \mu ^{2}]\, [\overline{m}^{2}_{\widetilde{\tau _{R}}}\, -\overline{m}_{\widetilde{\mu }_{L}}^{2}]}\, \times \, \sum ^{n=1}_{n=0}\, \frac{(-1)^{n}}{\overline{M}_{n}^{2}}\, \times \; \; \; \; \; \; \; \; \; \; \; \; \; \; \; \; \; \; \; \; \; \\
\; \; \; \; \; \; \; \; \; \; \; \; \; \; \; \; \left\{ A_{1b1}(X_{\widetilde{w}_{n}}\, ,\, X_{\widetilde{\mu }_{n}})\, +\, A_{1b2}(X_{\widetilde{w}_{n}}\, ,\, X_{\widetilde{\mu }_{n}})\right\} \, \overline{U}_{\nu _{\mu }}(p-q)\, P_{R}\, \sigma ^{\mu \nu }\, \varepsilon _{\mu }(q)\, q^{\nu }\, U_{\nu _{\tau }}(p)\\

\end{array}\]

+and finally for diagrams (2c):

\[
\begin{array}{c}
\\
\begin{array}{c}
\\
T_{2c}\, =\, \frac{i\, g_{2}^{2}\, e\, m_{\nu _{\tau }}\, \mu \, \tan (\beta )\, A_{e_{32}}\, }{16\, \pi ^{2}\, [M_{2}^{2}\, -\, \mu ^{2}]\, [\overline{m}^{2}_{\widetilde{\tau _{R}}}\, -\overline{m}_{\widetilde{\mu }_{L}}^{2}]}\, \times \, \sum ^{n=1}_{n=0}\, \frac{(-1)^{n}}{\overline{M}_{n}^{2}}\, \times \; \; \; \; \; \; \; \; \; \; \; \; \; \; \; \; \; \; \; \; \; \\
\; \; \; \; \; \; \; \; \; \; \; \; \; \; \; \; \left\{ A_{1c1}(X_{\widetilde{w}_{n}}\, ,\, X_{\widetilde{\mu }_{n}})\, +\, A_{1c2}(X_{\widetilde{w}_{n}}\, ,\, X_{\widetilde{\mu }_{n}})\right\} \, \overline{U}_{\nu _{\mu }}(p-q)\, P_{R}\, \sigma ^{\mu \nu }\, \varepsilon _{\mu }(q)\, q^{\nu }\, U_{\nu _{\tau }}(p)\\

\end{array}\\
\\

\end{array}\]

Then,due to these different expressions,we can calculate the supersymmetric
contributions to the transition magnetic moment of the neutrino in the MSSMRN
model.However,in this model there is no right-right off-diagonal components
\( (\overline{m}_{\widetilde{e}}^{2})_{ij} \).So,now we consider the \( SU(5)RN \)
model.

\subsection{amplitudes contributing in the \protect\( SU(5)RN\protect \) model}

As previously,with the same conditions,the left-left component (eq\ref{In1})
takes the same expression as in MSSMRN model.For left-right component \( A_{e_{32}} \)
we have an additional term dependent on Kobayashi-Maskawa matrix.Then,the left-right
off-diagonal component (only the dominant part) is: 
\begin{equation}
\label{In4}
A_{e_{32}}\, \simeq \, -\frac{3}{8\pi ^{2}}\, a_{0}\, [\, f_{e_{3}}\, V_{33}^{*}\, V_{32}\, f_{\nu _{3}}^{2}\, \log (\frac{M_{G}}{M_{\nu _{3}}})+3\, f_{e_{2}}\, V_{KM_{32}}^{*}\, V_{KM_{33}}\, f^{2}_{u_{3}}\, \log (\frac{M_{G}}{M_{GUT}})]
\end{equation}

\( f_{e_{i}} \) and \( f_{u_{3}} \) expressions are given in annex A.For the
right-right off-diagonal component we have:

\begin{equation}
\label{In5}
(m_{\widetilde{e}}^{2})_{32}\, \simeq \, -\frac{3}{8\pi ^{2}}\, (3\, m^{2}_{0}\, +\, a^{2}_{0})\, V_{KM_{33}}V^{*}_{KM_{32}}\, f^{2}_{u_{3}}\, \log (\frac{M_{G}}{M_{GUT}})
\end{equation}

So there are three kinds of diagrams contributing to the decay.The left-left
and left-right non-diagonal mass insertion diagrams are identical with MSSMRN
ones (figures \ref{f1} and \ref{f2}).Then,the amplitudes are the same.However,it
is necessary to replace \( A_{e_{32}} \) by its \( SU(5)RN \) expression (eq\ref{In4}).The
right-right non-diagonal mass insertion induces four diagrams drawn on figure(\ref{f3}).The
amplitudes associated to these diagrams are:

\( \begin{array}{c}
\\
T_{3a}\, =\, \frac{-\, i\, g_{2}^{2}\, e\, m_{\nu _{\tau }}\, M_{2}\, (\overline{m}_{LR}^{2})_{\mu }\, (\overline{m}_{\widetilde{e}}^{2})_{32}\, }{16\, \pi ^{2}\, [M_{2}^{2}\, -\, \mu ^{2}]\, [\overline{m}^{2}_{\widetilde{\tau _{R}}}\, -\overline{m}_{\widetilde{\mu }_{R}}^{2}]}\, \times \, \sum ^{n=3}_{n=0}\, \frac{(-1)^{n}}{[\overline{m}_{n}^{2}\, -\overline{m}_{\widetilde{\mu }_{L}}^{2}]}\, \frac{1}{\overline{M}_{n}^{2}}\, \times \; \; \; \; \; \; \; \; \; \; \; \; \; \; \; \; \; \; \; \; \; \\
\; \; \; \; \; \; \; \; \; \; \; \; \; \; \; \; \left\{ A_{1b1}(X_{\widetilde{w}_{n}}\, ,\, X_{\widetilde{\mu }_{n}})\, +\, A_{1b2}(X_{\widetilde{w}_{n}}\, ,\, X_{\widetilde{\mu }_{n}})\right\} \, \overline{U}_{\nu _{\mu }}(p-q)\, P_{R}\, \sigma ^{\mu \nu }\, \varepsilon _{\mu }(q)\, q^{\nu }\, U_{\nu _{\tau }}(p)\\

\end{array} \)

\( \begin{array}{c}
\\
T_{3b}\, =\, \frac{-\, i\, g_{2}^{2}\, e\, m_{\nu _{\tau }}\, \mu \: \tan (\beta )\, (\overline{m}_{LR}^{2})_{\mu }\, (\overline{m}_{\widetilde{e}}^{2})_{32}\, }{16\, \pi ^{2}\, [M_{2}^{2}\, -\, \mu ^{2}]\, [\overline{m}^{2}_{\widetilde{\tau _{R}}}\, -\overline{m}_{\widetilde{\mu }_{R}}^{2}]}\, \times \, \sum ^{n=3}_{n=0}\, \frac{(-1)^{n}}{[\overline{m}_{n}^{2}\, -\overline{m}_{\widetilde{\mu }_{L}}^{2}]}\, \frac{1}{\overline{M}_{n}^{2}}\, \times \; \; \; \; \; \; \; \; \; \; \; \; \; \; \; \; \; \; \; \; \; \\
\; \; \; \; \; \; \; \; \; \; \; \; \; \; \; \; \left\{ A_{1c1}(X_{\widetilde{w}_{n}}\, ,\, X_{\widetilde{\mu }_{n}})\, +\, A_{1c2}(X_{\widetilde{w}_{n}}\, ,\, X_{\widetilde{\mu }_{n}})\right\} \, \overline{U}_{\nu _{\mu }}(p-q)\, P_{R}\, \sigma ^{\mu \nu }\, \varepsilon _{\mu }(q)\, q^{\nu }\, U_{\nu _{\tau }}(p)\\

\end{array} \)

\begin{itemize}
\item 
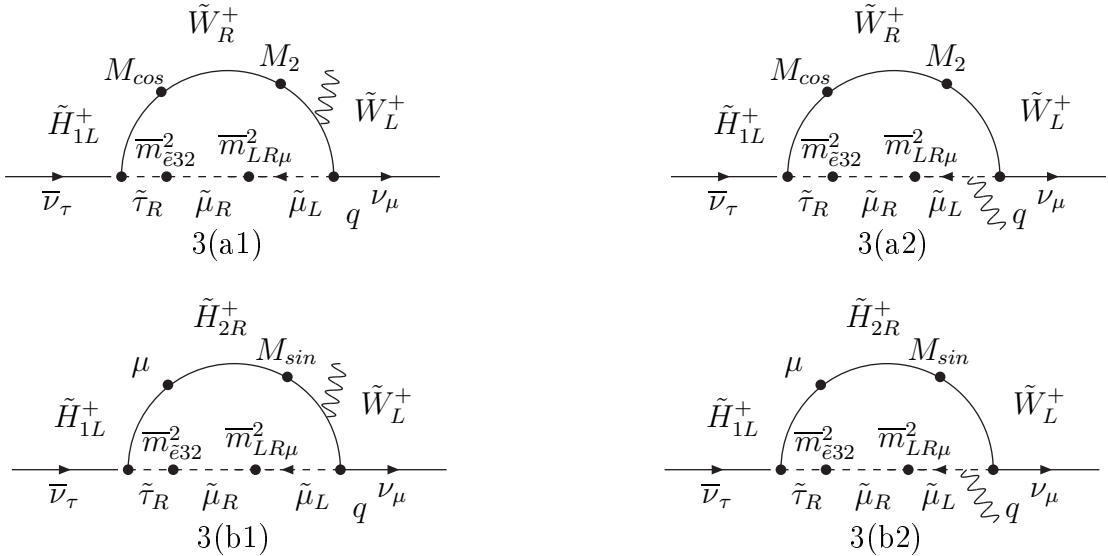
\begin{figure}
{\par\centering \begin{picture}(195,120)(0,0) \ArrowLine(0,20)(40,20)\Text(20,15)[tc]{\(\overline{\nu}_{\tau}\)}\Text(26,33)[b c]{\(\tilde{H}^{+}_{1L}\)}\par}

{\par\centering \Vertex(40,20){2}\par}

{\par\centering \DashLine(95,20)(40,20){3} \Text(50,15)[tc]{\(\tilde {\tau}_{R}\)} \Vertex(57,20){2}\par}

{\par\centering \Text(75,15)[tc]{\(\tilde {\mu}_{R}\)}\par}

{\par\centering \Vertex(88,20){2}\par}

{\par\centering \Text(91,24)[bc]{\(\overline{m}^{2}_{LR{\mu}}\)}\par}

{\par\centering \Text(80,0)[tc]{\ref{f3}(a1)}\Text(57,24)[bc]{\(\overline{m}^{2}_{\tilde{e}{32}}\)}\par}

{\par\centering \DashArrowLine(120,20)(80,20){3}\Text(110,15)[tc]{\(\tilde{\mu}_{L}\)}\par}

{\par\centering \Vertex(120,20){2}\par}

{\par\centering \ArrowLine(120,20)(160,20)\Text(140,15)[tc]{\({\nu}_{\mu}\)}\Text(138,38)[bc] {\(\tilde{W}^{+}_{L}\)}\par}

{\par\centering \CArc(80,20)(40,0,60)\par}

{\par\centering \Text(75,71)[bc]{\(\tilde{W}^{+}_{R}\)}\par}

{\par\centering \Vertex(100,55){2}\par}

{\par\centering \Text(100,60)[bc]{\({M}_{2}\)}\par}

{\par\centering \CArc(80,20)(40,60,180)\par}

{\par\centering \Vertex(55,52){2}\par}

{\par\centering \Text(45,55)[bc]{\(M_{cos}\)}\par}

{\par\centering \Photon(115,40)(120,60){-3}{4}\Text(125,0)[bl]{\(q\)}\par}

{\par\centering \end{picture}\hskip2cm \begin{picture}(170,80)(0,0) \ArrowLine(0,20)(40,20)\Text(20,15)[tc]{\(\overline{\nu}_{\tau}\)}\Text(26,33)[bc]{\(\tilde{H}^{+}_{1L}\)}\par}

{\par\centering \Vertex(40,20){2}\par}

{\par\centering \DashLine(95,20)(40,20){3} \Text(50,15)[tc]{\(\tilde {\tau}_{R}\)} \Vertex(57,20){2}\par}

{\par\centering \Text(75,15)[tc]{\(\tilde {\mu}_{R}\)}\par}

{\par\centering \Vertex(88,20){2}\par}

{\par\centering \Text(91,24)[bc]{\(\overline{m}^{2}_{LR{\mu}}\)}\par}

{\par\centering \Text(80,0)[tc]{\ref{f3}(a2)}\Text(57,24)[bc]{\(\overline{m}^{2}_{\tilde{e}{32}}\)}\par}

{\par\centering \DashArrowLine(120,20)(80,20){3}\Text(100,15)[tc]{\(\tilde{\mu}_{L}\)}\par}

{\par\centering \Vertex(120,20){2}\par}

{\par\centering \ArrowLine(120,20)(160,20)\Text(140,15)[tc]{\({\nu}_{\mu}\)}\Text(138,38)[bc]{\(\tilde{W}^{+}_{L}\)}\par}

{\par\centering \CArc(80,20)(40,0,60)\par}

{\par\centering \Text(75,71)[bc]{\(\tilde{W}^{+}_{R}\)}\par}

{\par\centering \Vertex(100,55){2}\par}

{\par\centering \Text(100,60)[bc]{\({M}_{2}\)}\par}

{\par\centering \CArc(80,20)(40,60,180)\par}

{\par\centering \Vertex(55,52){2}\par}

{\par\centering \Text(45,55)[bc]{\(M_{cos}\)}\par}

{\par\centering \Photon(110,20)(120,0){-3}{4}\Text(125,0)[bl]{\(q\)}\par}

{\par\centering \end{picture}\par}

{\par\centering \begin{picture}(190,110)(0,0) \ArrowLine(0,20)(40,20)\Text(20,15)[tc]{\(\overline{\nu}_{\tau}\)}\Text(26,33)[bc]{\(\tilde{H}^{+}_{1L}\)}\par}

{\par\centering \Vertex(40,20){2}\par}

{\par\centering \DashLine(95,20)(40,20){3} \Text(50,15)[tc]{\(\tilde {\tau}_{R}\)} \Vertex(57,20){2}\par}

{\par\centering \Text(75,15)[tc]{\(\tilde {\mu}_{R}\)}\par}

{\par\centering \Vertex(88,20){2}\par}

{\par\centering \Text(91,24)[bc]{\(\overline{m}^{2}_{LR{\mu}}\)}\par}

{\par\centering \Text(80,0)[tc]{\ref{f3}(b1)}\Text(57,24)[bc]{\(\overline{m}^{2}_{\tilde{e}{32}}\)}\par}

{\par\centering \DashArrowLine(120,20)(80,20){3}\Text(110,15)[tc]{\(\tilde{\mu}_{L}\)}\par}

{\par\centering \Vertex(120,20){2}\par}

{\par\centering \ArrowLine(120,20)(160,20)\Text(140,15)[tc]{\({\nu}_{\mu}\)}\Text(138,38)[bc]{\(\tilde{W}^{+}_{L}\)}\par}

{\par\centering \CArc(80,20)(40,0,60)\par}

{\par\centering \Text(75,71)[bc]{\(\tilde{H}^{+}_{2R}\)}\par}

{\par\centering \Vertex(100,55){2}\par}

{\par\centering \Text(100,60)[bc]{\({M}_{sin}\)}\par}

{\par\centering \CArc(80,20)(40,60,180)\par}

{\par\centering \Vertex(55,52){2}\par}

{\par\centering \Text(45,55)[bc]{\({\mu}\)}\par}

{\par\centering \Photon(115,40)(120,60){-3}{4}\Text(125,0)[bl]{\(q\)}\par}

{\par\centering \end{picture}\hskip2cm \begin{picture}(170,80)(0,0) \ArrowLine(0,20)(40,20)\Text(20,15)[tc]{\(\overline{\nu}_{\tau}\)}\Text(26,33)[bc]{\(\tilde{H}^{+}_{1L}\)}\par}

{\par\centering \Vertex(40,20){2}\par}

{\par\centering \DashLine(95,20)(40,20){3} \Text(50,15)[tc]{\(\tilde {\tau}_{R}\)} \Vertex(57,20){2}\par}

{\par\centering \Text(75,15)[tc]{\(\tilde {\mu}_{R}\)}\par}

{\par\centering \Vertex(88,20){2}\par}

{\par\centering \Text(91,24)[bc]{\(\overline{m}^{2}_{LR{\mu}}\)}\par}

{\par\centering \Text(80,0)[tc]{\ref{f3}(b2)}\Text(57,24)[bc]{\(\overline{m}^{2}_{\tilde{e}{32}}\)}\par}

{\par\centering \DashArrowLine(120,20)(80,20){3}\Text(100,15)[tc]{\(\tilde{\mu}_{L}\)}\par}

{\par\centering \Vertex(120,20){2}\par}

{\par\centering \ArrowLine(120,20)(160,20)\Text(140,15)[tc]{\({\nu}_{\mu}\)}\par}

{\par\centering \Text(138,38)[bc]{\(\tilde{W}^{+}_{L}\)}\par}

{\par\centering \CArc(80,20)(40,0,60)\par}

{\par\centering \Text(75,71)[bc]{\(\tilde{H}^{+}_{2R}\)}\par}

{\par\centering \Vertex(100,55){2}\par}

{\par\centering \Text(100,60)[bc]{\({M}_{sin}\)}\par}

{\par\centering \CArc(80,20)(40,60,180)\par}

{\par\centering \Vertex(55,52){2}\par}

{\par\centering \Text(45,55)[bc]{\({\mu}\)}\par}

{\par\centering \Photon(110,20)(120,0){-3}{4}\Text(125,0)[bl]{\(q\)}\par}

{\par\centering \end{picture}\par}

\caption{\label{f3}Chargino contributions with photon \protect\( (\gamma )\protect \)
and charged sleptons \protect\( (\, \widetilde{\tau },\widetilde{\mu }\, )\protect \)
coupling to right-right non-diagonal mass insertion :\protect\( (\overline{m}^{2}_{\widetilde{e}})_{32}\protect \)
.}
\end{figure}

\end{itemize}
\( \begin{array}{c}
\\
\; \; \; with\; \; \; \; \; \overline{m}_{n}^{2}\: =\, \left\{ \begin{array}{c}
\overline{m}_{\widetilde{\tau }_{R}}^{2}\; ,\; for\; n\, =0,1\\
\overline{m}_{\widetilde{\mu }_{R}}^{2}\; ,\; for\; n\, =2,3
\end{array}\right. \: \: \: \; \; ,\; \; \; \; \overline{M}_{n}^{2}\: =\, \left\{ \begin{array}{c}
\overline{m}_{\widetilde{\tau }_{R}}^{2}\; ,\; for\; n\, =0\\
\overline{m}_{\widetilde{\mu }_{R}}^{2}\; ,\; for\; n\, =3\\
\overline{m}_{\widetilde{\mu }_{L}}^{2}\; ,\; for\; n\, =1,2
\end{array}\right. \\
\\

\end{array} \)

\begin{equation}
\label{In6}
(\overline{m}_{LR}^{2})_{\mu }\, =\, m_{\mu }\, \mu \, \tan (\beta )\, -\, a_{0}\, m_{0}\, m_{\mu }\; \; ,\; m_{\mu }\; =\; lepton\; muon\; mass
\end{equation}
Therefore,in the next section we give the supersymmetric contributions to the
transition magnetic moment of the neutrino.

\section{Decay rate and magnetic moment in the MSSMRN and \protect\( SU(5)RN\protect \)
models}

The decay rate takes the form:\( \Gamma \, (\overline{\nu _{\tau }}\: \longrightarrow \: \nu _{\mu }+\gamma \, )\, =\, \frac{1}{3\pi }\, \frac{1}{m_{\nu _{\tau }}}\, \left| T\right| ^{2} \)
where T is the sum of the different amplitudes calculated in the previous section
for the two supersymmetric models.The neutrino magnetic moment is given as:\( K\, =\, \mu _{B}\, \tau _{sec}^{-\frac{1}{2}}\, m_{\nu _{\tau }}^{-\frac{3}{2}} \),
\( \mu _{B}\, =\, \frac{eh}{2m_{e}}\, \approx \, 0.44 \) is the Bohr magneton,\( \tau _{sec} \)
is the mean life and \( m_{\nu } \) the neutrino mass.The main aim is to compare
The MSSMRN and the \( SU(5)RN \) contributions to the neutrino magnetic moment
( \( K_{MSSMRN}\: \: \: ,\: \: \: K_{SU(5)RN} \) respectively ) with the SM
one.i.e. \( K_{SM}\, =\, \mu _{B}\, \times 3\times 10^{-19}\, m_{\nu }\, /eV \).From
the atmospheric result,we take the tau neutrino mass:\( m_{\nu }\, =\, 0.07eV\; \; and\; \; V_{32}\, =\, \frac{-1}{\sqrt{2}} \),then
we obtain \( K_{SM}\, \approx \, 9.24\times 10^{-21} \).For the right-handed
neutrino masses,we impose \( M_{\nu _{1}}\, =\, M_{\nu _{2}}\, =\, M_{\nu _{3}} \).For
our calculation,we take into account two different \( M_{\nu } \) values:\( M_{\nu }\, =\, 10^{12}GEV \)
and \( M_{\nu }\, =\, 10^{15}GEV \) and \( f_{u_{3}}\, =\, f_{\nu _{3}} \)
at the gravitational scale:\( M_{G}\, =\, 2\times 10^{18}GEV \).The GUT scale
is fixed by suspect2 program\footnote{%
kneur@lpm.univ-montp2.fr
}:\( M_{GUT}\, =\, 1.9\times 10^{16} \).Concerning SUSY parameters that is:
\( m_{0}\, \; ,\; m_{1/2}\; ,\; \tan (\beta )\; ,\; a_{0}\; ,\; sign(\mu ) \),we
choose four sets of values compatible with the relic density .These values were
obtained with an interface that relies darksusy and suspect2 programs.The \( a_{0} \)
parameter doesn't change significantly the sparticle mass spectrum for a set
of values,then \( a_{0} \) runs from zero to \( 500 \).For a given set,different
\( \tan (\beta ) \) values (\( \tan (\beta )\, =\, 10,20,30,40 \) ) don't
change significantly neutralino ( \( \chi ^{0}_{1} \) ) mass and then the relic
density ( \( \Omega ^{2} \) ) associated.In table{[}\ref{tab1}{]} we find
the sets of values and in table{[}\ref{tab2}{]}the mass spectrum associated
to each assignment for different \( \tan (\beta ) \) values.
\begin{table}
{\centering \begin{tabular}{|c|c|c|c|c|}
\hline 
&
set1&
set2&
set3&
set4\\
\hline 
\hline 
\( m_{0} \)&
186&
266&
360&
500\\
\hline 
\( m_{1/2} \)&
457&
290&
469&
597\\
\hline 
\( sign(\mu ) \)&
+&
+&
+&
+\\
\hline 
\( \tan (\beta ) \)&
10;20;30;40&
10;20;30;40&
10;20;30;40&
10;20;30;40\\
\hline 
\( a_{0} \)&
0;10;50;100;200;300;400;500&
`` ''&
`` ``&
`` ``\\
\hline 
\( \chi ^{0}_{1}mass \)&
186&
113&
191.8&
247.3\\
\hline 
\( \Omega ^{2} \)&
0.4&
0.7&
0.7&
0.02\\
\hline 
\end{tabular}\par}

\caption{\label{tab1}different SUSY parameters compatible with the relic density (mass
in GEV)}
\end{table}
\begin{table}
{\centering \begin{tabular}{|c|c|c|c|c|}
\hline 
&
set1&
set2&
set3&
set4\\
\hline 
\( M_{2} \)&
354&
222&
363&
465\\
\hline 
\( sm_{\widetilde{\mu }_{R}} \)&
255&
288&
401&
555\\
\hline 
\( sm_{\widetilde{\mu }_{L}} \)&
362.5&
332&
480&
648\\
\hline 
\( \tan (\beta )=10\, ,\, \mu  \)&
543&
346&
556&
695.5\\
\( sm_{\widetilde{\tau }_{R}} \)&
251.5&
286&
397.5&
543.5\\
\( sm_{\widetilde{\tau }_{L}} \)&
361&
331&
478.5&
640\\
\hline 
\( \tan (\beta )=20\, ,\, \mu  \)&
553&
346&
549&
686\\
\( sm_{\widetilde{\tau }_{R}} \)&
240.5&
276.5&
386&
528\\
\( sm_{\widetilde{\tau }_{L}} \)&
357.5&
327&
474&
633\\
\hline 
\( \tan (\beta )=30\, ,\, \mu  \)&
535.5&
346&
546&
681\\
\( sm_{\widetilde{\tau }_{R}} \)&
230&
262.5&
367&
502\\
\( sm_{\widetilde{\tau }_{L}} \)&
353.5&
321&
466&
623\\
\hline 
\( \tan (\beta )=40\, ,\, \mu  \)&
534.5&
346&
545&
680\\
\( sm_{\widetilde{\tau }_{R}} \)&
212.5&
242&
340.5&
466\\
\( sm_{\widetilde{\tau }_{L}} \)&
348&
313&
456&
609\\
\hline 
\end{tabular}\par}

\caption{\label{tab2}sparticle mass spectrum.The wino\protect\( (M_{2})\protect \)and
the sleptons mu (\protect\( sm_{\widetilde{\mu }_{R}}\protect \),\protect\( sm_{\widetilde{\mu }_{L}}\protect \))masses
are not significantly sensitive to \protect\( \tan (\beta )\protect \) .However,the
sleptons tau (\protect\( sm_{\widetilde{t}_{R}}\, ,\, sm_{\widetilde{t}_{L}}\protect \)
) and higgsino ( \protect\( \mu \protect \) ) masses are more sensitive to
\protect\( \tan (\beta )\protect \)}
\end{table}
Then,we have calculated the supersymmetric contributions to the neutrino magnetic
moment in MSSMRN and \( SU(5)RN \) models.

For \( a_{0}=0 \),in the MSSMRN model we have one insertion: \( (\overline{m}^{2})_{32} \),whereas
in the \( SU(5)RN \) case there are two insertions:\( (\overline{m}_{\widetilde{e}}^{2})_{32} \)
and \( (\overline{m}^{2})_{32} \).In such case, \( (\overline{m}_{LR}^{2})_{\tau }\,  \)and
\( (\overline{m}_{LR}^{2})_{\mu } \) take the follow form :\( (\overline{m}_{LR}^{2})_{\tau }\, =\, m_{\tau }\, \mu \, \tan (\beta )\,  \)and
\( (\overline{m}_{LR}^{2})_{\mu }\, =\, m_{\mu }\, \mu \, \tan (\beta )\,  \).The
results are presented on the figures(\ref{fi4}-\ref{fig5})
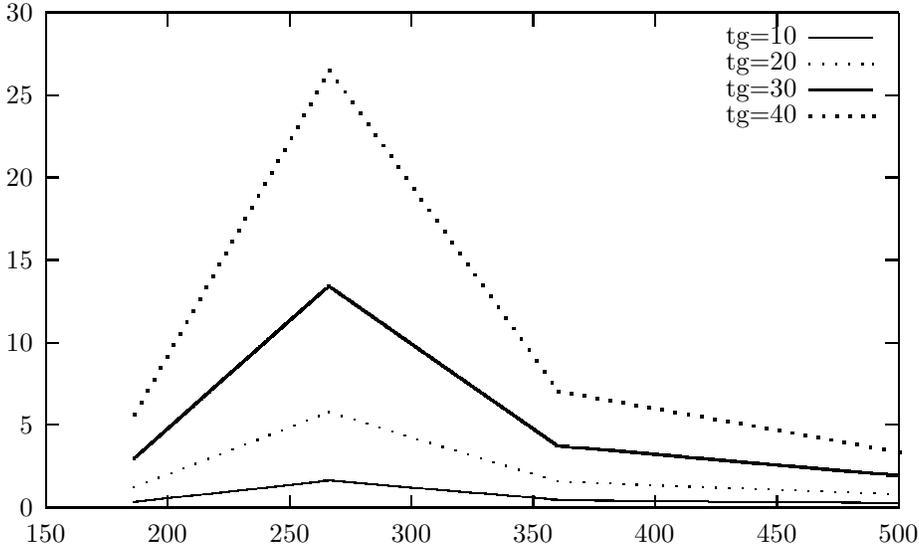
\begin{figure}
\setlength{\unitlength}{0.240900pt}
\ifx\plotpoint\undefined\newsavebox{\plotpoint}\fi
\sbox{\plotpoint}{\rule[-0.200pt]{0.400pt}{0.400pt}}%
\begin{picture}(1500,900)(0,0)
\font\gnuplot=cmr10 at 10pt
\gnuplot
\sbox{\plotpoint}{\rule[-0.200pt]{0.400pt}{0.400pt}}%
\put(100.0,82.0){\rule[-0.200pt]{4.818pt}{0.400pt}}
\put(80,82){\makebox(0,0)[r]{0}}
\put(1419.0,82.0){\rule[-0.200pt]{4.818pt}{0.400pt}}
\put(100.0,212.0){\rule[-0.200pt]{4.818pt}{0.400pt}}
\put(80,212){\makebox(0,0)[r]{5}}
\put(1419.0,212.0){\rule[-0.200pt]{4.818pt}{0.400pt}}
\put(100.0,341.0){\rule[-0.200pt]{4.818pt}{0.400pt}}
\put(80,341){\makebox(0,0)[r]{10}}
\put(1419.0,341.0){\rule[-0.200pt]{4.818pt}{0.400pt}}
\put(100.0,471.0){\rule[-0.200pt]{4.818pt}{0.400pt}}
\put(80,471){\makebox(0,0)[r]{15}}
\put(1419.0,471.0){\rule[-0.200pt]{4.818pt}{0.400pt}}
\put(100.0,601.0){\rule[-0.200pt]{4.818pt}{0.400pt}}
\put(80,601){\makebox(0,0)[r]{20}}
\put(1419.0,601.0){\rule[-0.200pt]{4.818pt}{0.400pt}}
\put(100.0,730.0){\rule[-0.200pt]{4.818pt}{0.400pt}}
\put(80,730){\makebox(0,0)[r]{25}}
\put(1419.0,730.0){\rule[-0.200pt]{4.818pt}{0.400pt}}
\put(100.0,860.0){\rule[-0.200pt]{4.818pt}{0.400pt}}
\put(80,860){\makebox(0,0)[r]{30}}
\put(1419.0,860.0){\rule[-0.200pt]{4.818pt}{0.400pt}}
\put(100.0,82.0){\rule[-0.200pt]{0.400pt}{4.818pt}}
\put(100,41){\makebox(0,0){150}}
\put(100.0,840.0){\rule[-0.200pt]{0.400pt}{4.818pt}}
\put(291.0,82.0){\rule[-0.200pt]{0.400pt}{4.818pt}}
\put(291,41){\makebox(0,0){200}}
\put(291.0,840.0){\rule[-0.200pt]{0.400pt}{4.818pt}}
\put(483.0,82.0){\rule[-0.200pt]{0.400pt}{4.818pt}}
\put(483,41){\makebox(0,0){250}}
\put(483.0,840.0){\rule[-0.200pt]{0.400pt}{4.818pt}}
\put(674.0,82.0){\rule[-0.200pt]{0.400pt}{4.818pt}}
\put(674,41){\makebox(0,0){300}}
\put(674.0,840.0){\rule[-0.200pt]{0.400pt}{4.818pt}}
\put(865.0,82.0){\rule[-0.200pt]{0.400pt}{4.818pt}}
\put(865,41){\makebox(0,0){350}}
\put(865.0,840.0){\rule[-0.200pt]{0.400pt}{4.818pt}}
\put(1056.0,82.0){\rule[-0.200pt]{0.400pt}{4.818pt}}
\put(1056,41){\makebox(0,0){400}}
\put(1056.0,840.0){\rule[-0.200pt]{0.400pt}{4.818pt}}
\put(1248.0,82.0){\rule[-0.200pt]{0.400pt}{4.818pt}}
\put(1248,41){\makebox(0,0){450}}
\put(1248.0,840.0){\rule[-0.200pt]{0.400pt}{4.818pt}}
\put(1439.0,82.0){\rule[-0.200pt]{0.400pt}{4.818pt}}
\put(1439,41){\makebox(0,0){500}}
\put(1439.0,840.0){\rule[-0.200pt]{0.400pt}{4.818pt}}
\put(100.0,82.0){\rule[-0.200pt]{322.565pt}{0.400pt}}
\put(1439.0,82.0){\rule[-0.200pt]{0.400pt}{187.420pt}}
\put(100.0,860.0){\rule[-0.200pt]{322.565pt}{0.400pt}}
\put(100.0,82.0){\rule[-0.200pt]{0.400pt}{187.420pt}}
\put(1279,820){\makebox(0,0)[r]{tg=10}}
\put(1299.0,820.0){\rule[-0.200pt]{24.090pt}{0.400pt}}
\put(238,91){\usebox{\plotpoint}}
\multiput(238.00,91.58)(4.541,0.498){65}{\rule{3.700pt}{0.120pt}}
\multiput(238.00,90.17)(298.320,34.000){2}{\rule{1.850pt}{0.400pt}}
\multiput(544.00,123.92)(5.851,-0.497){59}{\rule{4.732pt}{0.120pt}}
\multiput(544.00,124.17)(349.178,-31.000){2}{\rule{2.366pt}{0.400pt}}
\multiput(903.00,92.93)(59.598,-0.477){7}{\rule{42.980pt}{0.115pt}}
\multiput(903.00,93.17)(446.793,-5.000){2}{\rule{21.490pt}{0.400pt}}
\put(1279,779){\makebox(0,0)[r]{tg=20}}
\multiput(1299,779)(20.756,0.000){5}{\usebox{\plotpoint}}
\put(1399,779){\usebox{\plotpoint}}
\put(238,114){\usebox{\plotpoint}}
\multiput(238,114)(19.366,7.468){16}{\usebox{\plotpoint}}
\multiput(544,232)(19.860,-6.030){18}{\usebox{\plotpoint}}
\multiput(903,123)(20.741,-0.774){26}{\usebox{\plotpoint}}
\put(1439,103){\usebox{\plotpoint}}
\sbox{\plotpoint}{\rule[-0.400pt]{0.800pt}{0.800pt}}%
\put(1279,738){\makebox(0,0)[r]{tg=30}}
\put(1299.0,738.0){\rule[-0.400pt]{24.090pt}{0.800pt}}
\put(238,159){\usebox{\plotpoint}}
\multiput(238.00,160.41)(0.564,0.500){535}{\rule{1.103pt}{0.121pt}}
\multiput(238.00,157.34)(303.710,271.000){2}{\rule{0.552pt}{0.800pt}}
\multiput(544.00,428.09)(0.715,-0.500){495}{\rule{1.344pt}{0.121pt}}
\multiput(544.00,428.34)(356.210,-251.000){2}{\rule{0.672pt}{0.800pt}}
\multiput(903.00,177.09)(5.783,-0.502){87}{\rule{9.323pt}{0.121pt}}
\multiput(903.00,177.34)(516.649,-47.000){2}{\rule{4.662pt}{0.800pt}}
\sbox{\plotpoint}{\rule[-0.500pt]{1.000pt}{1.000pt}}%
\put(1279,697){\makebox(0,0)[r]{tg=40}}
\multiput(1299,697)(20.756,0.000){5}{\usebox{\plotpoint}}
\put(1399,697){\usebox{\plotpoint}}
\put(238,229){\usebox{\plotpoint}}
\multiput(238,229)(10.233,18.058){30}{\usebox{\plotpoint}}
\multiput(544,769)(12.042,-16.905){30}{\usebox{\plotpoint}}
\multiput(903,265)(20.430,-3.659){26}{\usebox{\plotpoint}}
\put(1439,169){\usebox{\plotpoint}}
\end{picture}

\caption{\label{fi4}neutrino magnetic moment in the MSSMRN model:\protect\( \frac{K_{MSSMRN}}{10^{-26}}(m_{0})\protect \)
, for \protect\( a_{0}=0\protect \) , \protect\( v=246GEV\protect \) and \protect\( M_{\nu }=10^{12}GEV\protect \)
.}
\end{figure}
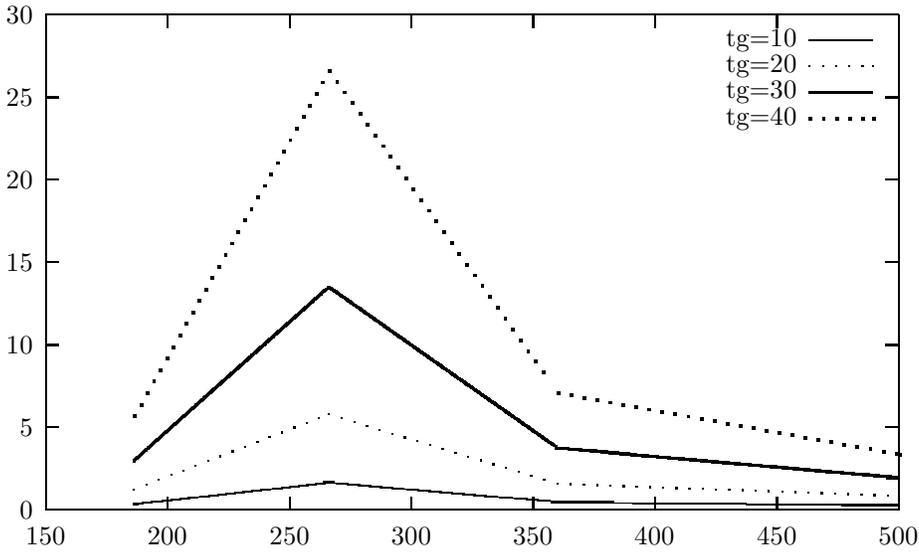
\begin{figure}
\setlength{\unitlength}{0.240900pt}
\ifx\plotpoint\undefined\newsavebox{\plotpoint}\fi
\sbox{\plotpoint}{\rule[-0.200pt]{0.400pt}{0.400pt}}%
\begin{picture}(1500,900)(0,0)
\font\gnuplot=cmr10 at 10pt
\gnuplot
\sbox{\plotpoint}{\rule[-0.200pt]{0.400pt}{0.400pt}}%
\put(100.0,82.0){\rule[-0.200pt]{4.818pt}{0.400pt}}
\put(80,82){\makebox(0,0)[r]{0}}
\put(1419.0,82.0){\rule[-0.200pt]{4.818pt}{0.400pt}}
\put(100.0,212.0){\rule[-0.200pt]{4.818pt}{0.400pt}}
\put(80,212){\makebox(0,0)[r]{5}}
\put(1419.0,212.0){\rule[-0.200pt]{4.818pt}{0.400pt}}
\put(100.0,341.0){\rule[-0.200pt]{4.818pt}{0.400pt}}
\put(80,341){\makebox(0,0)[r]{10}}
\put(1419.0,341.0){\rule[-0.200pt]{4.818pt}{0.400pt}}
\put(100.0,471.0){\rule[-0.200pt]{4.818pt}{0.400pt}}
\put(80,471){\makebox(0,0)[r]{15}}
\put(1419.0,471.0){\rule[-0.200pt]{4.818pt}{0.400pt}}
\put(100.0,601.0){\rule[-0.200pt]{4.818pt}{0.400pt}}
\put(80,601){\makebox(0,0)[r]{20}}
\put(1419.0,601.0){\rule[-0.200pt]{4.818pt}{0.400pt}}
\put(100.0,730.0){\rule[-0.200pt]{4.818pt}{0.400pt}}
\put(80,730){\makebox(0,0)[r]{25}}
\put(1419.0,730.0){\rule[-0.200pt]{4.818pt}{0.400pt}}
\put(100.0,860.0){\rule[-0.200pt]{4.818pt}{0.400pt}}
\put(80,860){\makebox(0,0)[r]{30}}
\put(1419.0,860.0){\rule[-0.200pt]{4.818pt}{0.400pt}}
\put(100.0,82.0){\rule[-0.200pt]{0.400pt}{4.818pt}}
\put(100,41){\makebox(0,0){150}}
\put(100.0,840.0){\rule[-0.200pt]{0.400pt}{4.818pt}}
\put(291.0,82.0){\rule[-0.200pt]{0.400pt}{4.818pt}}
\put(291,41){\makebox(0,0){200}}
\put(291.0,840.0){\rule[-0.200pt]{0.400pt}{4.818pt}}
\put(483.0,82.0){\rule[-0.200pt]{0.400pt}{4.818pt}}
\put(483,41){\makebox(0,0){250}}
\put(483.0,840.0){\rule[-0.200pt]{0.400pt}{4.818pt}}
\put(674.0,82.0){\rule[-0.200pt]{0.400pt}{4.818pt}}
\put(674,41){\makebox(0,0){300}}
\put(674.0,840.0){\rule[-0.200pt]{0.400pt}{4.818pt}}
\put(865.0,82.0){\rule[-0.200pt]{0.400pt}{4.818pt}}
\put(865,41){\makebox(0,0){350}}
\put(865.0,840.0){\rule[-0.200pt]{0.400pt}{4.818pt}}
\put(1056.0,82.0){\rule[-0.200pt]{0.400pt}{4.818pt}}
\put(1056,41){\makebox(0,0){400}}
\put(1056.0,840.0){\rule[-0.200pt]{0.400pt}{4.818pt}}
\put(1248.0,82.0){\rule[-0.200pt]{0.400pt}{4.818pt}}
\put(1248,41){\makebox(0,0){450}}
\put(1248.0,840.0){\rule[-0.200pt]{0.400pt}{4.818pt}}
\put(1439.0,82.0){\rule[-0.200pt]{0.400pt}{4.818pt}}
\put(1439,41){\makebox(0,0){500}}
\put(1439.0,840.0){\rule[-0.200pt]{0.400pt}{4.818pt}}
\put(100.0,82.0){\rule[-0.200pt]{322.565pt}{0.400pt}}
\put(1439.0,82.0){\rule[-0.200pt]{0.400pt}{187.420pt}}
\put(100.0,860.0){\rule[-0.200pt]{322.565pt}{0.400pt}}
\put(100.0,82.0){\rule[-0.200pt]{0.400pt}{187.420pt}}
\put(1279,820){\makebox(0,0)[r]{tg=10}}
\put(1299.0,820.0){\rule[-0.200pt]{24.090pt}{0.400pt}}
\put(238,91){\usebox{\plotpoint}}
\multiput(238.00,91.58)(4.541,0.498){65}{\rule{3.700pt}{0.120pt}}
\multiput(238.00,90.17)(298.320,34.000){2}{\rule{1.850pt}{0.400pt}}
\multiput(544.00,123.92)(5.851,-0.497){59}{\rule{4.732pt}{0.120pt}}
\multiput(544.00,124.17)(349.178,-31.000){2}{\rule{2.366pt}{0.400pt}}
\multiput(903.00,92.93)(59.598,-0.477){7}{\rule{42.980pt}{0.115pt}}
\multiput(903.00,93.17)(446.793,-5.000){2}{\rule{21.490pt}{0.400pt}}
\put(1279,779){\makebox(0,0)[r]{tg=20}}
\multiput(1299,779)(20.756,0.000){5}{\usebox{\plotpoint}}
\put(1399,779){\usebox{\plotpoint}}
\put(238,114){\usebox{\plotpoint}}
\multiput(238,114)(19.366,7.468){16}{\usebox{\plotpoint}}
\multiput(544,232)(19.860,-6.030){18}{\usebox{\plotpoint}}
\multiput(903,123)(20.742,-0.735){26}{\usebox{\plotpoint}}
\put(1439,104){\usebox{\plotpoint}}
\sbox{\plotpoint}{\rule[-0.400pt]{0.800pt}{0.800pt}}%
\put(1279,738){\makebox(0,0)[r]{tg=30}}
\put(1299.0,738.0){\rule[-0.400pt]{24.090pt}{0.800pt}}
\put(238,159){\usebox{\plotpoint}}
\multiput(238.00,160.41)(0.560,0.500){539}{\rule{1.097pt}{0.121pt}}
\multiput(238.00,157.34)(303.724,273.000){2}{\rule{0.548pt}{0.800pt}}
\multiput(544.00,430.09)(0.710,-0.500){499}{\rule{1.335pt}{0.121pt}}
\multiput(544.00,430.34)(356.229,-253.000){2}{\rule{0.668pt}{0.800pt}}
\multiput(903.00,177.09)(5.783,-0.502){87}{\rule{9.323pt}{0.121pt}}
\multiput(903.00,177.34)(516.649,-47.000){2}{\rule{4.662pt}{0.800pt}}
\sbox{\plotpoint}{\rule[-0.500pt]{1.000pt}{1.000pt}}%
\put(1279,697){\makebox(0,0)[r]{tg=40}}
\multiput(1299,697)(20.756,0.000){5}{\usebox{\plotpoint}}
\put(1399,697){\usebox{\plotpoint}}
\put(238,229){\usebox{\plotpoint}}
\multiput(238,229)(10.190,18.082){31}{\usebox{\plotpoint}}
\multiput(544,772)(11.994,-16.939){29}{\usebox{\plotpoint}}
\multiput(903,265)(20.430,-3.659){27}{\usebox{\plotpoint}}
\put(1439,169){\usebox{\plotpoint}}
\end{picture}

\caption{\label{fig5}neutrino magnetic moment in the \protect\( SU(5)_{RN}\protect \)
model:\protect\( \frac{K_{SU(5)RN}}{10^{-26}}(m_{0})\protect \), for \protect\( a_{0}=0\protect \),\protect\( v=246GEV\protect \),\protect\( M_{\nu }=10^{12}GEV\protect \)
and \protect\( (V_{KM})_{32}=-0.035\protect \).}
\end{figure}
In the \( SU(5)_{RN} \) case,we take \( (V_{KM})_{32}=-0.035 \).The results
don't change significantly for \( (V_{KM})_{32}=-0.043 \).

On figures (\ref{fi4}-\ref{fig5}) we remark that \( K_{MSSMRN} \) and \( K_{SU(5)RN} \)
are smaller than \( K_{SM}\simeq 9.24\times 10^{-21} \)(\( K_{MSSMRN}\simeq 2.65\times 10^{-25}\; ,K_{SU(5)RN}\simeq 2.66\times 10^{-25}\; for\, \tan (\beta )=40 \)
the greatest case for \( a_{0}=0 \)).However \( K_{SU(5)RN} \) is ,approximately
,\( 0.6 \) percent greater than \( K_{MSSMRN} \).Otherwise,the various results
are reduced by a factor \( 0.5 \) for \( M_{\nu }=10^{15}GEV \) .Nevertheless,plots
fig.(\ref{fi4}-\ref{fig5}) show a highest value for \( m_{O}=266GEV \).Then
we have retained the set2 for SUSY parameters.So for the set2,we have taken
into account different \( a_{0} \) values and we present the results on figures
(\ref{fig6}-\ref{fig7}-\ref{fig8}).When \( a_{0} \) has a value different
of zero the left-right insertion \( (A_{e})_{32} \) contributes and \( \overline{m}_{LR}^{2} \)
eq(\ref{In2},\ref{In6})is very sensitive to the \( a_{0} \) parameter.
\begin{figure}
\setlength{\unitlength}{0.240900pt}
\ifx\plotpoint\undefined\newsavebox{\plotpoint}\fi
\sbox{\plotpoint}{\rule[-0.200pt]{0.400pt}{0.400pt}}%
\begin{picture}(1500,900)(0,0)
\font\gnuplot=cmr10 at 10pt
\gnuplot
\sbox{\plotpoint}{\rule[-0.200pt]{0.400pt}{0.400pt}}%
\put(100.0,82.0){\rule[-0.200pt]{4.818pt}{0.400pt}}
\put(80,82){\makebox(0,0)[r]{0}}
\put(1419.0,82.0){\rule[-0.200pt]{4.818pt}{0.400pt}}
\put(100.0,168.0){\rule[-0.200pt]{4.818pt}{0.400pt}}
\put(80,168){\makebox(0,0)[r]{2}}
\put(1419.0,168.0){\rule[-0.200pt]{4.818pt}{0.400pt}}
\put(100.0,255.0){\rule[-0.200pt]{4.818pt}{0.400pt}}
\put(80,255){\makebox(0,0)[r]{4}}
\put(1419.0,255.0){\rule[-0.200pt]{4.818pt}{0.400pt}}
\put(100.0,341.0){\rule[-0.200pt]{4.818pt}{0.400pt}}
\put(80,341){\makebox(0,0)[r]{6}}
\put(1419.0,341.0){\rule[-0.200pt]{4.818pt}{0.400pt}}
\put(100.0,428.0){\rule[-0.200pt]{4.818pt}{0.400pt}}
\put(80,428){\makebox(0,0)[r]{8}}
\put(1419.0,428.0){\rule[-0.200pt]{4.818pt}{0.400pt}}
\put(100.0,514.0){\rule[-0.200pt]{4.818pt}{0.400pt}}
\put(80,514){\makebox(0,0)[r]{10}}
\put(1419.0,514.0){\rule[-0.200pt]{4.818pt}{0.400pt}}
\put(100.0,601.0){\rule[-0.200pt]{4.818pt}{0.400pt}}
\put(80,601){\makebox(0,0)[r]{12}}
\put(1419.0,601.0){\rule[-0.200pt]{4.818pt}{0.400pt}}
\put(100.0,687.0){\rule[-0.200pt]{4.818pt}{0.400pt}}
\put(80,687){\makebox(0,0)[r]{14}}
\put(1419.0,687.0){\rule[-0.200pt]{4.818pt}{0.400pt}}
\put(100.0,774.0){\rule[-0.200pt]{4.818pt}{0.400pt}}
\put(80,774){\makebox(0,0)[r]{16}}
\put(1419.0,774.0){\rule[-0.200pt]{4.818pt}{0.400pt}}
\put(100.0,860.0){\rule[-0.200pt]{4.818pt}{0.400pt}}
\put(80,860){\makebox(0,0)[r]{18}}
\put(1419.0,860.0){\rule[-0.200pt]{4.818pt}{0.400pt}}
\put(100.0,82.0){\rule[-0.200pt]{0.400pt}{4.818pt}}
\put(100,41){\makebox(0,0){0}}
\put(100.0,840.0){\rule[-0.200pt]{0.400pt}{4.818pt}}
\put(234.0,82.0){\rule[-0.200pt]{0.400pt}{4.818pt}}
\put(234,41){\makebox(0,0){50}}
\put(234.0,840.0){\rule[-0.200pt]{0.400pt}{4.818pt}}
\put(368.0,82.0){\rule[-0.200pt]{0.400pt}{4.818pt}}
\put(368,41){\makebox(0,0){100}}
\put(368.0,840.0){\rule[-0.200pt]{0.400pt}{4.818pt}}
\put(502.0,82.0){\rule[-0.200pt]{0.400pt}{4.818pt}}
\put(502,41){\makebox(0,0){150}}
\put(502.0,840.0){\rule[-0.200pt]{0.400pt}{4.818pt}}
\put(636.0,82.0){\rule[-0.200pt]{0.400pt}{4.818pt}}
\put(636,41){\makebox(0,0){200}}
\put(636.0,840.0){\rule[-0.200pt]{0.400pt}{4.818pt}}
\put(769.0,82.0){\rule[-0.200pt]{0.400pt}{4.818pt}}
\put(769,41){\makebox(0,0){250}}
\put(769.0,840.0){\rule[-0.200pt]{0.400pt}{4.818pt}}
\put(903.0,82.0){\rule[-0.200pt]{0.400pt}{4.818pt}}
\put(903,41){\makebox(0,0){300}}
\put(903.0,840.0){\rule[-0.200pt]{0.400pt}{4.818pt}}
\put(1037.0,82.0){\rule[-0.200pt]{0.400pt}{4.818pt}}
\put(1037,41){\makebox(0,0){350}}
\put(1037.0,840.0){\rule[-0.200pt]{0.400pt}{4.818pt}}
\put(1171.0,82.0){\rule[-0.200pt]{0.400pt}{4.818pt}}
\put(1171,41){\makebox(0,0){400}}
\put(1171.0,840.0){\rule[-0.200pt]{0.400pt}{4.818pt}}
\put(1305.0,82.0){\rule[-0.200pt]{0.400pt}{4.818pt}}
\put(1305,41){\makebox(0,0){450}}
\put(1305.0,840.0){\rule[-0.200pt]{0.400pt}{4.818pt}}
\put(1439.0,82.0){\rule[-0.200pt]{0.400pt}{4.818pt}}
\put(1439,41){\makebox(0,0){500}}
\put(1439.0,840.0){\rule[-0.200pt]{0.400pt}{4.818pt}}
\put(100.0,82.0){\rule[-0.200pt]{322.565pt}{0.400pt}}
\put(1439.0,82.0){\rule[-0.200pt]{0.400pt}{187.420pt}}
\put(100.0,860.0){\rule[-0.200pt]{322.565pt}{0.400pt}}
\put(100.0,82.0){\rule[-0.200pt]{0.400pt}{187.420pt}}
\put(1279,820){\makebox(0,0)[r]{tg=10}}
\put(1299.0,820.0){\rule[-0.200pt]{24.090pt}{0.400pt}}
\put(127,82){\usebox{\plotpoint}}
\multiput(127.00,82.61)(23.681,0.447){3}{\rule{14.367pt}{0.108pt}}
\multiput(127.00,81.17)(77.181,3.000){2}{\rule{7.183pt}{0.400pt}}
\multiput(234.00,85.59)(7.776,0.489){15}{\rule{6.056pt}{0.118pt}}
\multiput(234.00,84.17)(121.431,9.000){2}{\rule{3.028pt}{0.400pt}}
\multiput(368.00,94.58)(3.212,0.498){81}{\rule{2.652pt}{0.120pt}}
\multiput(368.00,93.17)(262.495,42.000){2}{\rule{1.326pt}{0.400pt}}
\multiput(636.00,136.58)(1.941,0.499){135}{\rule{1.648pt}{0.120pt}}
\multiput(636.00,135.17)(263.580,69.000){2}{\rule{0.824pt}{0.400pt}}
\multiput(903.00,205.58)(1.356,0.499){195}{\rule{1.183pt}{0.120pt}}
\multiput(903.00,204.17)(265.545,99.000){2}{\rule{0.591pt}{0.400pt}}
\multiput(1171.00,304.58)(1.056,0.499){251}{\rule{0.944pt}{0.120pt}}
\multiput(1171.00,303.17)(266.040,127.000){2}{\rule{0.472pt}{0.400pt}}
\put(1279,779){\makebox(0,0)[r]{tg=20}}
\multiput(1299,779)(20.756,0.000){5}{\usebox{\plotpoint}}
\put(1399,779){\usebox{\plotpoint}}
\put(127,82){\usebox{\plotpoint}}
\multiput(127,82)(20.741,0.775){6}{\usebox{\plotpoint}}
\multiput(234,86)(20.571,2.763){6}{\usebox{\plotpoint}}
\multiput(368,104)(19.868,6.005){14}{\usebox{\plotpoint}}
\multiput(636,185)(18.382,9.638){14}{\usebox{\plotpoint}}
\multiput(903,325)(16.694,12.333){16}{\usebox{\plotpoint}}
\multiput(1171,523)(14.925,14.424){18}{\usebox{\plotpoint}}
\put(1439,782){\usebox{\plotpoint}}
\end{picture}

\caption{\label{fig6}\protect\( \frac{K_{MMSSMRN}}{10^{-20}}(a_{0})\protect \) for
\protect\( m_{0}=266GEV\protect \) (set2) \protect\( M_{\nu }=10^{12}GEV\protect \)
(for \protect\( M_{\nu }=10^{15}GEV\protect \) there is a decrease of 0.5 )}
\end{figure}
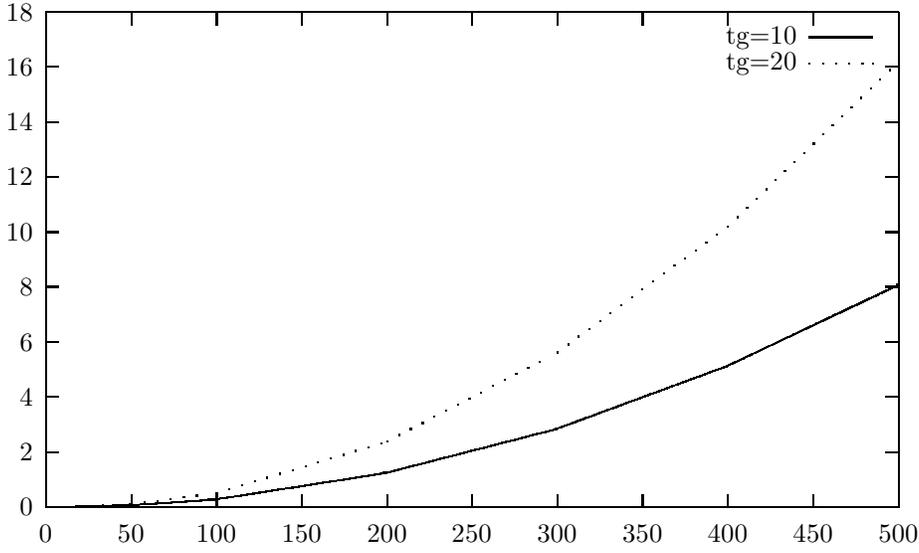
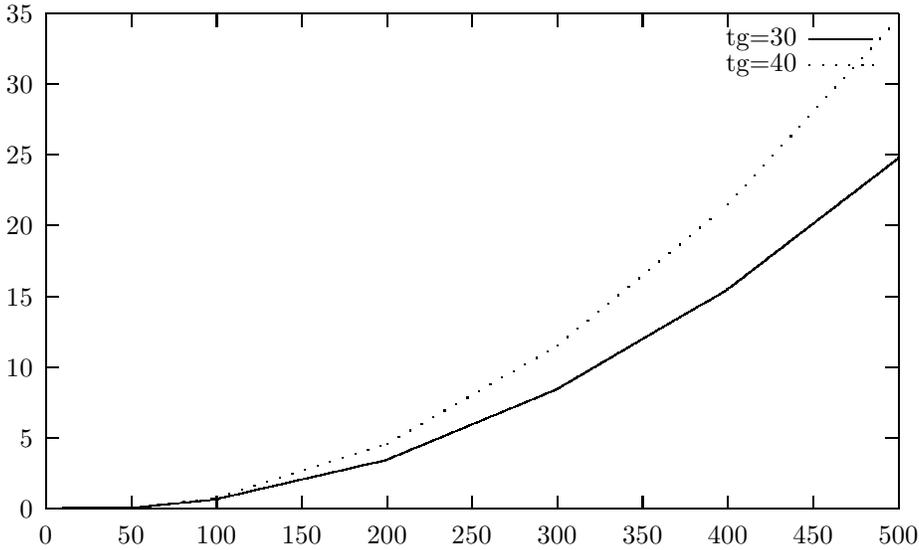
\begin{figure}
\setlength{\unitlength}{0.240900pt}
\ifx\plotpoint\undefined\newsavebox{\plotpoint}\fi
\begin{picture}(1500,900)(0,0)
\font\gnuplot=cmr10 at 10pt
\gnuplot
\sbox{\plotpoint}{\rule[-0.200pt]{0.400pt}{0.400pt}}%
\put(100.0,82.0){\rule[-0.200pt]{4.818pt}{0.400pt}}
\put(80,82){\makebox(0,0)[r]{0}}
\put(1419.0,82.0){\rule[-0.200pt]{4.818pt}{0.400pt}}
\put(100.0,193.0){\rule[-0.200pt]{4.818pt}{0.400pt}}
\put(80,193){\makebox(0,0)[r]{5}}
\put(1419.0,193.0){\rule[-0.200pt]{4.818pt}{0.400pt}}
\put(100.0,304.0){\rule[-0.200pt]{4.818pt}{0.400pt}}
\put(80,304){\makebox(0,0)[r]{10}}
\put(1419.0,304.0){\rule[-0.200pt]{4.818pt}{0.400pt}}
\put(100.0,415.0){\rule[-0.200pt]{4.818pt}{0.400pt}}
\put(80,415){\makebox(0,0)[r]{15}}
\put(1419.0,415.0){\rule[-0.200pt]{4.818pt}{0.400pt}}
\put(100.0,527.0){\rule[-0.200pt]{4.818pt}{0.400pt}}
\put(80,527){\makebox(0,0)[r]{20}}
\put(1419.0,527.0){\rule[-0.200pt]{4.818pt}{0.400pt}}
\put(100.0,638.0){\rule[-0.200pt]{4.818pt}{0.400pt}}
\put(80,638){\makebox(0,0)[r]{25}}
\put(1419.0,638.0){\rule[-0.200pt]{4.818pt}{0.400pt}}
\put(100.0,749.0){\rule[-0.200pt]{4.818pt}{0.400pt}}
\put(80,749){\makebox(0,0)[r]{30}}
\put(1419.0,749.0){\rule[-0.200pt]{4.818pt}{0.400pt}}
\put(100.0,860.0){\rule[-0.200pt]{4.818pt}{0.400pt}}
\put(80,860){\makebox(0,0)[r]{35}}
\put(1419.0,860.0){\rule[-0.200pt]{4.818pt}{0.400pt}}
\put(100.0,82.0){\rule[-0.200pt]{0.400pt}{4.818pt}}
\put(100,41){\makebox(0,0){0}}
\put(100.0,840.0){\rule[-0.200pt]{0.400pt}{4.818pt}}
\put(234.0,82.0){\rule[-0.200pt]{0.400pt}{4.818pt}}
\put(234,41){\makebox(0,0){50}}
\put(234.0,840.0){\rule[-0.200pt]{0.400pt}{4.818pt}}
\put(368.0,82.0){\rule[-0.200pt]{0.400pt}{4.818pt}}
\put(368,41){\makebox(0,0){100}}
\put(368.0,840.0){\rule[-0.200pt]{0.400pt}{4.818pt}}
\put(502.0,82.0){\rule[-0.200pt]{0.400pt}{4.818pt}}
\put(502,41){\makebox(0,0){150}}
\put(502.0,840.0){\rule[-0.200pt]{0.400pt}{4.818pt}}
\put(636.0,82.0){\rule[-0.200pt]{0.400pt}{4.818pt}}
\put(636,41){\makebox(0,0){200}}
\put(636.0,840.0){\rule[-0.200pt]{0.400pt}{4.818pt}}
\put(769.0,82.0){\rule[-0.200pt]{0.400pt}{4.818pt}}
\put(769,41){\makebox(0,0){250}}
\put(769.0,840.0){\rule[-0.200pt]{0.400pt}{4.818pt}}
\put(903.0,82.0){\rule[-0.200pt]{0.400pt}{4.818pt}}
\put(903,41){\makebox(0,0){300}}
\put(903.0,840.0){\rule[-0.200pt]{0.400pt}{4.818pt}}
\put(1037.0,82.0){\rule[-0.200pt]{0.400pt}{4.818pt}}
\put(1037,41){\makebox(0,0){350}}
\put(1037.0,840.0){\rule[-0.200pt]{0.400pt}{4.818pt}}
\put(1171.0,82.0){\rule[-0.200pt]{0.400pt}{4.818pt}}
\put(1171,41){\makebox(0,0){400}}
\put(1171.0,840.0){\rule[-0.200pt]{0.400pt}{4.818pt}}
\put(1305.0,82.0){\rule[-0.200pt]{0.400pt}{4.818pt}}
\put(1305,41){\makebox(0,0){450}}
\put(1305.0,840.0){\rule[-0.200pt]{0.400pt}{4.818pt}}
\put(1439.0,82.0){\rule[-0.200pt]{0.400pt}{4.818pt}}
\put(1439,41){\makebox(0,0){500}}
\put(1439.0,840.0){\rule[-0.200pt]{0.400pt}{4.818pt}}
\put(100.0,82.0){\rule[-0.200pt]{322.565pt}{0.400pt}}
\put(1439.0,82.0){\rule[-0.200pt]{0.400pt}{187.420pt}}
\put(100.0,860.0){\rule[-0.200pt]{322.565pt}{0.400pt}}
\put(100.0,82.0){\rule[-0.200pt]{0.400pt}{187.420pt}}
\put(1279,820){\makebox(0,0)[r]{tg=30}}
\put(1299.0,820.0){\rule[-0.200pt]{24.090pt}{0.400pt}}
\put(127,83){\usebox{\plotpoint}}
\multiput(234.00,83.58)(4.904,0.494){25}{\rule{3.929pt}{0.119pt}}
\multiput(234.00,82.17)(125.846,14.000){2}{\rule{1.964pt}{0.400pt}}
\multiput(368.00,97.58)(2.170,0.499){121}{\rule{1.829pt}{0.120pt}}
\multiput(368.00,96.17)(264.204,62.000){2}{\rule{0.915pt}{0.400pt}}
\multiput(636.00,159.58)(1.205,0.499){219}{\rule{1.062pt}{0.120pt}}
\multiput(636.00,158.17)(264.795,111.000){2}{\rule{0.531pt}{0.400pt}}
\multiput(903.00,270.58)(0.854,0.499){311}{\rule{0.783pt}{0.120pt}}
\multiput(903.00,269.17)(266.375,157.000){2}{\rule{0.391pt}{0.400pt}}
\multiput(1171.00,427.58)(0.651,0.500){409}{\rule{0.620pt}{0.120pt}}
\multiput(1171.00,426.17)(266.712,206.000){2}{\rule{0.310pt}{0.400pt}}
\put(127.0,83.0){\rule[-0.200pt]{25.776pt}{0.400pt}}
\put(1279,779){\makebox(0,0)[r]{tg=40}}
\multiput(1299,779)(20.756,0.000){5}{\usebox{\plotpoint}}
\put(1399,779){\usebox{\plotpoint}}
\put(127,83){\usebox{\plotpoint}}
\multiput(127,83)(20.755,-0.194){6}{\usebox{\plotpoint}}
\multiput(234,82)(20.590,2.612){6}{\usebox{\plotpoint}}
\multiput(368,99)(19.784,6.275){14}{\usebox{\plotpoint}}
\multiput(636,184)(17.979,10.370){15}{\usebox{\plotpoint}}
\multiput(903,338)(15.984,13.240){16}{\usebox{\plotpoint}}
\multiput(1171,560)(14.008,15.315){19}{\usebox{\plotpoint}}
\put(1439,853){\usebox{\plotpoint}}
\end{picture}

\caption{\label{fig7}\protect\( \frac{K_{MSSMRN}}{10^{-20}}(a_{0})\protect \) for
\protect\( m_{0}=266GEV\protect \) (set2) \protect\( M_{\nu }=10^{12}GEV\protect \)(
for \protect\( M_{\nu }=10^{15}GEV\protect \) there is a decrease of 0.5 )}
\end{figure}
\begin{figure}

\setlength{\unitlength}{0.240900pt}
\ifx\plotpoint\undefined\newsavebox{\plotpoint}\fi
\begin{picture}(1500,900)(0,0)
\font\gnuplot=cmr10 at 10pt
\gnuplot
\sbox{\plotpoint}{\rule[-0.200pt]{0.400pt}{0.400pt}}%
\put(100.0,82.0){\rule[-0.200pt]{4.818pt}{0.400pt}}
\put(80,82){\makebox(0,0)[r]{0}}
\put(1419.0,82.0){\rule[-0.200pt]{4.818pt}{0.400pt}}
\put(100.0,168.0){\rule[-0.200pt]{4.818pt}{0.400pt}}
\put(80,168){\makebox(0,0)[r]{2}}
\put(1419.0,168.0){\rule[-0.200pt]{4.818pt}{0.400pt}}
\put(100.0,255.0){\rule[-0.200pt]{4.818pt}{0.400pt}}
\put(80,255){\makebox(0,0)[r]{4}}
\put(1419.0,255.0){\rule[-0.200pt]{4.818pt}{0.400pt}}
\put(100.0,341.0){\rule[-0.200pt]{4.818pt}{0.400pt}}
\put(80,341){\makebox(0,0)[r]{6}}
\put(1419.0,341.0){\rule[-0.200pt]{4.818pt}{0.400pt}}
\put(100.0,428.0){\rule[-0.200pt]{4.818pt}{0.400pt}}
\put(80,428){\makebox(0,0)[r]{8}}
\put(1419.0,428.0){\rule[-0.200pt]{4.818pt}{0.400pt}}
\put(100.0,514.0){\rule[-0.200pt]{4.818pt}{0.400pt}}
\put(80,514){\makebox(0,0)[r]{10}}
\put(1419.0,514.0){\rule[-0.200pt]{4.818pt}{0.400pt}}
\put(100.0,601.0){\rule[-0.200pt]{4.818pt}{0.400pt}}
\put(80,601){\makebox(0,0)[r]{12}}
\put(1419.0,601.0){\rule[-0.200pt]{4.818pt}{0.400pt}}
\put(100.0,687.0){\rule[-0.200pt]{4.818pt}{0.400pt}}
\put(80,687){\makebox(0,0)[r]{14}}
\put(1419.0,687.0){\rule[-0.200pt]{4.818pt}{0.400pt}}
\put(100.0,774.0){\rule[-0.200pt]{4.818pt}{0.400pt}}
\put(80,774){\makebox(0,0)[r]{16}}
\put(1419.0,774.0){\rule[-0.200pt]{4.818pt}{0.400pt}}
\put(100.0,860.0){\rule[-0.200pt]{4.818pt}{0.400pt}}
\put(80,860){\makebox(0,0)[r]{18}}
\put(1419.0,860.0){\rule[-0.200pt]{4.818pt}{0.400pt}}
\put(100.0,82.0){\rule[-0.200pt]{0.400pt}{4.818pt}}
\put(100,41){\makebox(0,0){0}}
\put(100.0,840.0){\rule[-0.200pt]{0.400pt}{4.818pt}}
\put(234.0,82.0){\rule[-0.200pt]{0.400pt}{4.818pt}}
\put(234,41){\makebox(0,0){50}}
\put(234.0,840.0){\rule[-0.200pt]{0.400pt}{4.818pt}}
\put(368.0,82.0){\rule[-0.200pt]{0.400pt}{4.818pt}}
\put(368,41){\makebox(0,0){100}}
\put(368.0,840.0){\rule[-0.200pt]{0.400pt}{4.818pt}}
\put(502.0,82.0){\rule[-0.200pt]{0.400pt}{4.818pt}}
\put(502,41){\makebox(0,0){150}}
\put(502.0,840.0){\rule[-0.200pt]{0.400pt}{4.818pt}}
\put(636.0,82.0){\rule[-0.200pt]{0.400pt}{4.818pt}}
\put(636,41){\makebox(0,0){200}}
\put(636.0,840.0){\rule[-0.200pt]{0.400pt}{4.818pt}}
\put(769.0,82.0){\rule[-0.200pt]{0.400pt}{4.818pt}}
\put(769,41){\makebox(0,0){250}}
\put(769.0,840.0){\rule[-0.200pt]{0.400pt}{4.818pt}}
\put(903.0,82.0){\rule[-0.200pt]{0.400pt}{4.818pt}}
\put(903,41){\makebox(0,0){300}}
\put(903.0,840.0){\rule[-0.200pt]{0.400pt}{4.818pt}}
\put(1037.0,82.0){\rule[-0.200pt]{0.400pt}{4.818pt}}
\put(1037,41){\makebox(0,0){350}}
\put(1037.0,840.0){\rule[-0.200pt]{0.400pt}{4.818pt}}
\put(1171.0,82.0){\rule[-0.200pt]{0.400pt}{4.818pt}}
\put(1171,41){\makebox(0,0){400}}
\put(1171.0,840.0){\rule[-0.200pt]{0.400pt}{4.818pt}}
\put(1305.0,82.0){\rule[-0.200pt]{0.400pt}{4.818pt}}
\put(1305,41){\makebox(0,0){450}}
\put(1305.0,840.0){\rule[-0.200pt]{0.400pt}{4.818pt}}
\put(1439.0,82.0){\rule[-0.200pt]{0.400pt}{4.818pt}}
\put(1439,41){\makebox(0,0){500}}
\put(1439.0,840.0){\rule[-0.200pt]{0.400pt}{4.818pt}}
\put(100.0,82.0){\rule[-0.200pt]{322.565pt}{0.400pt}}
\put(1439.0,82.0){\rule[-0.200pt]{0.400pt}{187.420pt}}
\put(100.0,860.0){\rule[-0.200pt]{322.565pt}{0.400pt}}
\put(100.0,82.0){\rule[-0.200pt]{0.400pt}{187.420pt}}
\put(1279,820){\makebox(0,0)[r]{tg=10}}
\put(1299.0,820.0){\rule[-0.200pt]{24.090pt}{0.400pt}}
\put(127,82){\usebox{\plotpoint}}
\multiput(127.00,82.61)(23.681,0.447){3}{\rule{14.367pt}{0.108pt}}
\multiput(127.00,81.17)(77.181,3.000){2}{\rule{7.183pt}{0.400pt}}
\multiput(234.00,85.59)(7.776,0.489){15}{\rule{6.056pt}{0.118pt}}
\multiput(234.00,84.17)(121.431,9.000){2}{\rule{3.028pt}{0.400pt}}
\multiput(368.00,94.58)(3.212,0.498){81}{\rule{2.652pt}{0.120pt}}
\multiput(368.00,93.17)(262.495,42.000){2}{\rule{1.326pt}{0.400pt}}
\multiput(636.00,136.58)(1.914,0.499){137}{\rule{1.626pt}{0.120pt}}
\multiput(636.00,135.17)(263.626,70.000){2}{\rule{0.813pt}{0.400pt}}
\multiput(903.00,206.58)(1.356,0.499){195}{\rule{1.183pt}{0.120pt}}
\multiput(903.00,205.17)(265.545,99.000){2}{\rule{0.591pt}{0.400pt}}
\multiput(1171.00,305.58)(1.056,0.499){251}{\rule{0.944pt}{0.120pt}}
\multiput(1171.00,304.17)(266.040,127.000){2}{\rule{0.472pt}{0.400pt}}
\put(1279,779){\makebox(0,0)[r]{tg=20}}
\multiput(1299,779)(20.756,0.000){5}{\usebox{\plotpoint}}
\put(1399,779){\usebox{\plotpoint}}
\put(127,82){\usebox{\plotpoint}}
\multiput(127,82)(20.741,0.775){6}{\usebox{\plotpoint}}
\multiput(234,86)(20.571,2.763){6}{\usebox{\plotpoint}}
\multiput(368,104)(19.868,6.005){14}{\usebox{\plotpoint}}
\multiput(636,185)(18.353,9.692){14}{\usebox{\plotpoint}}
\multiput(903,326)(16.723,12.293){16}{\usebox{\plotpoint}}
\multiput(1171,523)(14.869,14.481){18}{\usebox{\plotpoint}}
\put(1439,784){\usebox{\plotpoint}}
\end{picture}

\caption{\label{fig8}\protect\( \frac{K_{SU(5)_{RN}}}{10^{-20}}(a_{0})\protect \)
for \protect\( m_{0}=266GEV\protect \)( set2)\protect\( M_{\nu }=10^{12}GEV\protect \)
( for \protect\( M_{\nu }=10^{15}GEV\protect \) there is a decrease of 0.5
)}
\end{figure}
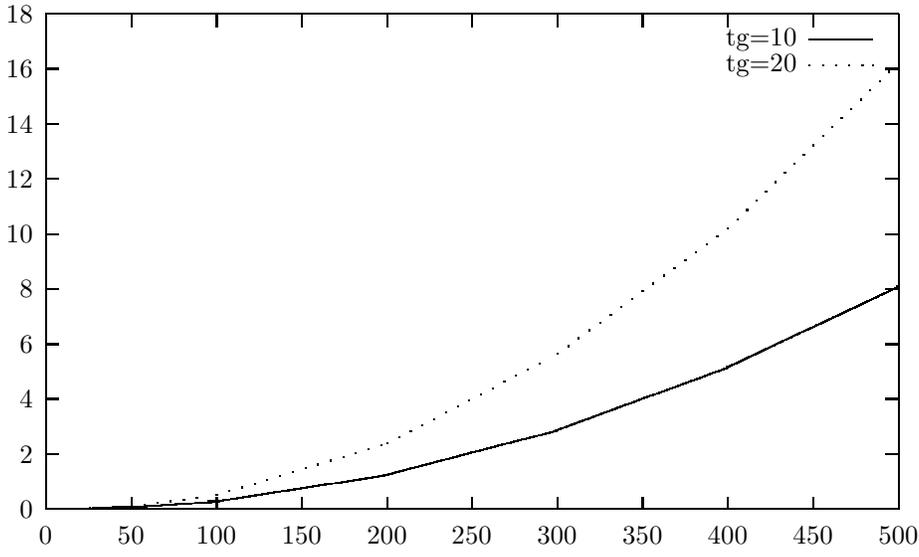
From figures (\ref{fig6},\ref{fig8}),we remark that \( K_{MSSMRN} \) and
\( K_{SU(5)RN} \) are the same order (\( \approx 0.6percent \) ) and very
sensitive to \( a_{0} \) and \( \tan (\beta ) \) values.Indeed,\( K_{MSSMRN} \)
and \( K_{SU(5)RN} \) increase when \( a_{0} \) and \( \tan (\beta ) \) increase:

+for \( a_{0}\leq 200 \) \( K_{MSSMRN} \) and \( K_{SU(5)RN} \),are a few
less than \( K_{SM} \) ,that is \( K=\frac{K_{MSSMRN}}{K_{SM}}\approx \frac{K_{SU(5)RN}}{K_{SM}}\approx 0.3\; for\; \tan (\beta )=10 \),
\( K=\frac{K_{MSSMRN}}{K_{SM}}\approx \frac{K_{SU(5)RN}}{K_{SM}}\approx 0.8\; for\; \tan (\beta )=40 \).

+ for \( a_{0}=200 \) ,\( K_{MSSMRN} \) and \( K_{SU(5)RN} \) ,are a few
times greater than \( \textrm{K}_{\textrm{SM}} \) : \( K=1.34\; for\; \tan (\beta )=10 \),\( K=4.95\; for\; \tan (\beta )=40 \)

+for \( a_{0}\geq 200 \) and more precisely \( a_{0}=500 \),we obtain \( K=8.7\; for\; \tan (\beta )=10,K=17.5\; for\; \tan (\beta )=20,K=27\; for\; \tan (\beta )=30,K=37.5\; for\; \tan (\beta )=40 \)
.

So,in this last case,the SUSY contributions \( K_{MSSMRN} \) and \( K_{SU(5)RN}\simeq 3.4\times 10^{-19} \)
are greater than \( K_{SM}\simeq 9.24\times 10^{-21} \).Nevertheless,SUSY contributions
(with right singlet neutrino \( M_{\nu }=10^{12}GEV \) ) to the transition
magnetic moment of the neutrino are very small compare to the experimental data
\cite{4}(for example: \( K_{Raffelt}\leq 10^{-12} \) , \( K_{Kolb}\leq 10^{-6} \)
, \( K_{Biller}\leq 10^{-9} \)).

\section{Conclusion}

In the context of the minimal N=1 Supergravity model,we have calculated the
transition magnetic moment of the neutrino in the MSSMRN and \( SU(5)RN \)
models.The results are the same order,for the two models,and very sensitive
to \( a_{0} \) and \( \tan (\beta ) \) parameters.For large \( a_{0} \) and
\( \tan (\beta ) \) values,the SUSY contributions are greater than SM ones
but remain weak compared with experimental data .

\section{Annex A}

+The interaction Lagrangian,in the \( MSSMRN \) model,of lepton-slepton and
wino (higgsino) components of chargino is:

\[
L_{int}\; =\; i\, (\, -g_{2}\, P_{R}\, \overline{\nu }_{I}\, \widetilde{W}^{+}_{L}\, \widetilde{l}_{IL}\: +\: g_{2}\, \frac{m_{\nu _{I}}}{\sqrt{2}\, m_{w}\, \cos (\beta )}\, P_{L}\, \overline{\nu }_{I}\, \widetilde{H}^{+}_{1L}\, \widetilde{l}_{IR}\, )\]

with:I=1-3 (index of generation), \( P_{R}\; \; and\; \; P_{L}\; chirality\; \; projectors \).\( m_{\nu }\;  \)is
the neutrino mass and \( \widetilde{w}\; \; and\; \; \widetilde{H}\; the \)
wino and higgsino component of chargino.\( \widetilde{l} \) stands for slepton.These
interactions are presented on figure \ref{feyn}

\begin{itemize}
\item 
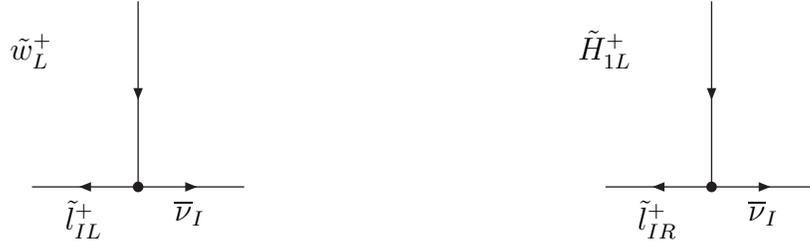
\begin{figure}
{\par\centering \begin{picture}(160,80)(0,0) 
\ArrowLine(120,20)(80,20)\Text(100,15)[tc]{\(\tilde {l}^{+}_{IL}\)}
\Vertex(120,20){2}
\ArrowLine(120,20)(160,20)\Text(140,15)[tc]{\(\overline{\nu}_{I}\)}
 \ArrowLine(120,90)(120,20) \Text(80,65)[bc]{\(\tilde{w}^{+}_{L}\)} \end{picture}\hskip2cm\begin{picture}(160,80)(0,0)\par}

{\par\centering \ArrowLine(120,20)(80,20)\Text(100,15)[tc]{\(\tilde{l}^{+}_{IR}\)}\par}

{\par\centering \Vertex(120,20){2}\par}

{\par\centering \ArrowLine(120,20)(160,20)\Text(140,15)[tc]{\(\overline{\nu}_{I}\)}\par}

{\par\centering \ArrowLine(120,90)(120,20) \Text(80,65)[bc]{\(\tilde{H}^{+}_{1L}\)} \end{picture}\par}

\caption{\label{feyn}Feynman rule }
\end{figure}
+ The Yukawa couplings are given by:
\item \( f_{e_{IJ}}=\, -\sqrt{2}\, \frac{m_{e_{I}}}{v\, \cos (\beta )}\, \delta _{IJ} \)
, \( f_{u_{IJ}}=\sqrt{2}\, \frac{m_{u_{I}}}{v\, \sin (\beta )}\, (V_{KM})_{IJ} \).Where
\( (V_{KM})_{IJ} \) is the Kobayashi-Maskawa matrix and \( m_{e_{I}} \) (\( m_{u_{I}} \))the
lepton mass ( quark-up mass).
\end{itemize}

\section{Annex B}

Now,we give analytic expressions of the Feynman integrals which appear in the
evaluation of the amplitudes.

\( \begin{array}{c}
\\
F_{a}(X_{\widetilde{W}_{n}},X_{\widetilde{\mu }_{n}})\, =\, \int ^{1}_{0}\, dx_{1}\, \int ^{x_{1}}_{0}\, dx_{2}\, \frac{1}{\left[ X_{\widetilde{W}_{n}}\, (1-x_{1})\, +\, X_{\widetilde{\mu }_{n}}\, (x_{1}-x_{2})\, +x_{2}\right] }\\

\end{array} \), 

\( \begin{array}{c}
\\
F_{b}(X)\, =\, \int ^{1}_{0}\, dx_{1}\, \int ^{x_{1}}_{0}\, dx_{2}\, \int ^{x_{2}}_{0}\, dx_{3}\, \frac{x_{2}-1}{\left[ X\, (1-x_{2})\, +\, x_{2}\right] ^{2}}\\

\end{array} \), 

\( \begin{array}{c}
\\
F_{c}(X)\, =\, \int ^{1}_{0}\, dx_{1}\, \int ^{x_{1}}_{0}\, dx_{2}\, \frac{x_{1}-1}{\left[ X\, (1-x_{1})\, +\, x_{1}\right] }\\

\end{array} \),

\( \begin{array}{c}
\\
G_{a}(X_{\widetilde{W}_{n}},X_{\widetilde{\mu }_{n}})\, =\, \int ^{1}_{0}\, dx_{1}\, \int ^{x_{1}}_{0}\, dx_{2}\, \frac{x_{1}}{\left[ X_{\widetilde{W}_{n}}\, (1-x_{1})\, +\, X_{\widetilde{\mu }_{n}}\, (x_{1}-x_{2})\, +x_{2}\right] }\\

\end{array} \),

\( \begin{array}{c}
\\
G_{b_{1}}(X_{\widetilde{W}_{n}},X_{\widetilde{\mu }_{n}})\, =\, \int ^{1}_{0}\, dx_{1}\, \int ^{x_{1}}_{0}\, dx_{2}\, \int ^{x_{2}}_{0}\, dx_{3}\, \frac{x_{1}}{\left[ X_{\widetilde{W}_{n}}\, (1-x_{1})\, +\, X_{\widetilde{\mu }_{n}}\, (x_{1}-x_{3})\, +x_{3}\right] ^{2}}\\

\end{array} \),

\( \begin{array}{c}
\\
G_{b_{2}}(X_{\widetilde{W}_{n}},X_{\widetilde{\mu }_{n}})\, =\, \int ^{1}_{0}\, dx_{1}\, \int ^{x_{1}}_{0}\, dx_{2}\, \int ^{x_{2}}_{0}\, dx_{3}\, \frac{1-x_{1}}{\left[ X_{\widetilde{W}_{n}}\, (1-x_{1})\, +\, X_{\widetilde{\mu }_{n}}\, (x_{1}-x_{3})\, +x_{3}\right] ^{2}}\\

\end{array} \),

\( \begin{array}{c}
\\
G_{b_{3}}(X_{\widetilde{W}_{n}},X_{\widetilde{\mu }_{n}})\, =\, \int ^{1}_{0}\, dx_{1}\, \int ^{x_{1}}_{0}\, dx_{2}\, \int ^{x_{2}}_{0}\, dx_{3}\, \frac{x_{1}}{\left[ X_{\widetilde{W}_{n}}\, (1-x_{1})\, +\, X_{\widetilde{\mu }_{n}}\, (x_{1}-x_{3})\, +x_{3}\right] }\\

\end{array} \),

\( \begin{array}{c}
\\
G_{b_{4}}(X_{\widetilde{W}_{n}},X_{\widetilde{\mu }_{n}})\, =\, \int ^{1}_{0}\, dx_{1}\, \int ^{x_{1}}_{0}\, dx_{2}\, \int ^{x_{2}}_{0}\, dx_{3}\, \frac{1-x_{1}}{\left[ X_{\widetilde{W}_{n}}\, (1-x_{1})\, +\, X_{\widetilde{\mu }_{n}}\, (x_{1}-x_{3})\, +x_{3}\right] }\\

\end{array} \),

\( \begin{array}{c}
\\
G_{c_{1}}(X)\, =\, \int ^{1}_{0}\, dx_{1}\, \int ^{x_{1}}_{0}\, dx_{2}\, \int ^{x_{2}}_{0}\, dx_{3}\, \frac{1}{\left[ X\, (1-x_{3})\, +\, x_{3}\right] ^{2}}\\

\end{array} \),

\( \begin{array}{c}
\\
G_{c_{2}}(X)\, =\, \int ^{1}_{0}\, dx_{1}\, \int ^{x_{1}}_{0}\, dx_{2}\, \int ^{x_{2}}_{0}\, dx_{3}\, \frac{x_{1}}{\left[ X\, (1-x_{3})\, +\, x_{3}\right] ^{2}}\\

\end{array} \),

\( \begin{array}{c}
\\
G_{c_{3}}(X)\, =\, \int ^{1}_{0}\, dx_{1}\, \int ^{x_{1}}_{0}\, dx_{2}\, \int ^{x_{2}}_{0}\, dx_{3}\, \frac{1}{\left[ X\, (1-x_{3})\, +\, x_{3}\right] }\\

\end{array} \),

\( \begin{array}{c}
\\
G_{c_{4}}(X)\, =\, \int ^{1}_{0}\, dx_{1}\, \int ^{x_{1}}_{0}\, dx_{2}\, \int ^{x_{2}}_{0}\, dx_{3}\, \frac{x_{1}}{\left[ X\, (1-x_{3})\, +\, x_{3}\right] }\\

\end{array} \),

\( \begin{array}{c}
\\
G_{d_{1}}(X)\, =\, \int ^{1}_{0}\, dx_{1}\, \int ^{x_{1}}_{0}\, dx_{2}\, \int ^{x_{2}}_{0}\, dx_{3}\, \frac{x_{2}}{\left[ X\, (1-x_{3})\, +\, x_{3}\right] ^{2}}\\

\end{array} \),

\( \begin{array}{c}
\\
G_{d_{2}}(X)\, =\, \int ^{1}_{0}\, dx_{1}\, \int ^{x_{1}}_{0}\, dx_{2}\, \int ^{x_{2}}_{0}\, dx_{3}\, \frac{x_{2}}{\left[ X\, (1-x_{3})\, +\, x_{3}\right] }\\

\end{array} \),

\( \begin{array}{c}
\\
G_{e}(X)\, =\, \int ^{1}_{0}\, dx_{1}\, \int ^{x_{1}}_{0}\, dx_{2}\, \int ^{x_{2}}_{0}\, dx_{3}\, \frac{x_{2}-1}{\left[ X\, (1-x_{2})\, +\, x_{2}\right] }\\

\end{array} \),

\( \begin{array}{c}
\\
H_{a}(X)\, =\, \int ^{1}_{0}\, dx_{1}\, \int ^{x_{1}}_{0}\, dx_{2}\frac{1}{\left[ X\, (1-x_{2})\, +\, x_{2}\right] }\\

\end{array} \),

\( \begin{array}{c}
\\
H_{b}(X_{\widetilde{W}_{n}},X_{\widetilde{\mu }_{n}})\, =\, \int ^{1}_{0}\, dx_{1}\, \int ^{x_{1}}_{0}\, dx_{2}\, \int ^{x_{2}}_{0}\, dx_{3}\, \frac{1}{\left[ X_{\widetilde{W}_{n}}\, (1-x_{2})\, +\, X_{\widetilde{\mu }_{n}}\, (x_{2}-x_{3})\, +x_{3}\right] }\\

\end{array} \),

\( \begin{array}{c}
\\
H_{c}(X_{\widetilde{W}_{n}},X_{\widetilde{\mu }_{n}})\, =\, \int ^{1}_{0}\, dx_{1}\, \int ^{x_{1}}_{0}\, dx_{2}\, \int ^{x_{2}}_{0}\, dx_{3}\, \frac{x_{2}}{\left[ X_{\widetilde{W}_{n}}\, (1-x_{2})\, +\, X_{\widetilde{\mu }_{n}}\, (x_{2}-x_{3})\, +x_{3}\right] }\\

\end{array} \),

\( \begin{array}{c}
\\
H_{d}(X_{\widetilde{W}_{n}},X_{\widetilde{\mu }_{n}})\, =\, \int ^{1}_{0}\, dx_{1}\, \int ^{x_{1}}_{0}\, dx_{2}\, \int ^{x_{2}}_{0}\, dx_{3}\, \frac{x_{2}}{\left[ X_{\widetilde{W}_{n}}\, (1-x_{2})\, +\, X_{\widetilde{\mu }_{n}}\, (x_{2}-x_{3})\, +x_{3}\right] ^{2}}\\

\end{array} \),

\( \begin{array}{c}
\\
H_{e}(X_{\widetilde{W}_{n}},X_{\widetilde{\mu }_{n}})\, =\, \int ^{1}_{0}\, dx_{1}\, \int ^{x_{1}}_{0}\, dx_{2}\, \int ^{x_{2}}_{0}\, dx_{3}\, \frac{1}{\left[ X_{\widetilde{W}_{n}}\, (1-x_{2})\, +\, X_{\widetilde{\mu }_{n}}\, (x_{2}-x_{3})\, +x_{3}\right] ^{2}}\\

\end{array} \),

\( \begin{array}{c}
\\
I(X)\, =\, \int ^{1}_{0}\, dx_{1}\, \int ^{x_{1}}_{0}\, dx_{2}\frac{x_{1}}{\left[ X\, (1-x_{2})\, +\, x_{2}\right] }\\

\end{array} \),

\( \begin{array}{c}
\\
h(X)\, =\, \int ^{1}_{0}\, dx_{1}\, \int ^{x_{1}}_{0}\, dx_{2}\frac{(x_{1}\, -x_{2}-1)\times (x_{1}-x_{2})}{\left[ 1+X\, (x_{1}-x_{2})\, +\, x_{2}-x_{1}\right] }\\

\end{array} \),

\( \begin{array}{c}
\\
f(X)\, =\, \int ^{1}_{0}\, dx_{1}\, \int ^{x_{1}}_{0}\, dx_{2}\frac{(x_{2}-1)x_{2}}{\left[ X\, (1-x_{2})\, +\, x_{2}\right] }\; =\; -\, 2\times g(X)\\

\end{array} \),

\( with\; \; \; X_{\widetilde{W}_{n}}\, =\, \frac{M_{2}^{2}}{\overline{M}^{2}_{n}}\; ,\; X_{\widetilde{\mu }_{n}}\, =\, \frac{\mu ^{2}}{\overline{M}_{n}^{2}}\; \; (\, M_{2}\: and\: \mu \; for\; wino\; and\; higgsino\; mass\, ) \).\( X \)
can be either \( X_{\widetilde{W}_{n}} \) or \( X_{\widetilde{\mu }_{n}} \).

+The expressions of \( A_{1b1}\; ,\; A_{1b_{2}}\; ,\; A_{1c_{1}}\; ,\; A_{1c_{2}} \)
terms are:

\( \begin{array}{c}
\\
A_{1b_{1}}(X_{\widetilde{W}_{n}},X_{\widetilde{\mu }_{n}})=\frac{1}{[M^{2}_{2}-\mu ^{2}]}\{-(\mu M^{2}_{2}+M^{3}_{2})F_{a}(X_{\widetilde{W}_{n}},X_{\widetilde{\mu }_{n}})+(2\mu ^{2}M_{2}+\mu M^{2}_{2}+M^{3}_{2})G_{a}(X_{\widetilde{W}_{n}},X_{\widetilde{\mu }_{n}})\\
+(2\mu M^{2}_{2}+M^{3}_{2})H_{a}(X_{\widetilde{W}_{n}})-(\mu M^{2}_{2}+M^{3}_{2})I(X_{\widetilde{W}_{n}})-\mu M^{2}_{2}H_{a}(X_{\widetilde{\mu }_{n}})-2\mu ^{2}M_{2}I(X_{\widetilde{\mu }_{n}})\}\\
+\frac{2}{\overline{M}^{2}_{n}}\{(2\mu ^{2}M_{2}-\mu M^{2}_{2}+M^{3}_{2})H_{d}(X_{\widetilde{W}_{n}},X_{\widetilde{\mu }_{n}})-M^{3}_{2}H_{e}(X_{\widetilde{W}_{n}},X_{\widetilde{\mu }_{n}})+(3\mu M^{2}_{2}+2M^{3}_{2})G_{c_{1}}(X_{\widetilde{W}_{n}})\\
-(\mu M^{2}_{2}+M^{3}_{2})G_{c_{2}}(X_{\widetilde{W}_{n}})-(M^{3}_{2}-\mu M^{2}_{2})G_{d_{1}}(X_{\widetilde{W}_{n}})\}\\
+3\times \{-M_{2}H_{b}(X_{\widetilde{W}_{n}},X_{\widetilde{\mu }_{n}})+(3M_{2}-\mu )H_{c}(X_{\widetilde{W}_{n}},X_{\widetilde{\mu }_{n}})-(\mu +2M_{2})G_{c_{3}}(X_{\widetilde{W}_{n}})-(\mu +M_{2})G_{c_{4}}(X_{\widetilde{W}_{n}})\\
+(\mu -M_{2})G_{d_{2}}(X_{\widetilde{W}_{n}})\}\\

\end{array} \)

\( A_{1b_{2}}(X_{\widetilde{W}_{n}},X_{\widetilde{\mu }_{n}})\, =\, \frac{\mu M^{2}_{2}}{[M^{2}_{2}\, -\, \mu ^{2}]}[\, F_{c_{2}}(X_{\widetilde{\mu }_{n}})\, -\, F_{c_{2}}(X_{\widetilde{W}_{n}})\, ]\, -\, \frac{2\mu M^{2}_{2}}{\overline{M}^{2}_{n}}\, F_{b}(X_{\widetilde{W}_{n}})\, +\, 3(\mu \, +\, 2M_{2})\, G_{e}(X_{\widetilde{W}_{n}}) \)

\( \begin{array}{c}
\\
A_{1c_{1}}(X_{\widetilde{W}_{n}},X_{\widetilde{\mu }_{n}})=\frac{1}{[M^{2}_{2}-\mu ^{2}]}\{2\mu M^{2}_{2}F_{a}(X_{\widetilde{W}_{n}},X_{\widetilde{\mu }_{n}})-(\mu ^{2}M_{2}+2\mu M^{2}_{2}+\mu ^{3})G_{a}(X_{\widetilde{W}_{n}},X_{\widetilde{\mu }_{n}})\\
-(\mu ^{2}M_{2}+2\mu M^{2}_{2})H_{a}(X_{\widetilde{W}_{n}})+2\mu M^{2}_{2}I(X_{\widetilde{W}_{n}})+\mu M^{2}_{2}H_{a}(X_{\widetilde{\mu }_{n}})+(\mu ^{2}M_{2}+\mu ^{3})I(X_{\widetilde{\mu }_{n}})\}\\
+\frac{2}{\overline{M}^{2}_{n}}\{\mu ^{3}G_{b_{1}}(X_{\widetilde{W}_{n}},X_{\widetilde{\mu }_{n}})+(\mu ^{2}M_{2}-2\mu M^{2}_{2})G_{b_{2}}(X_{\widetilde{W}_{n}},X_{\widetilde{\mu }_{n}})+(\mu ^{2}M_{2}-\mu ^{3})G_{c_{2}}(X_{\widetilde{\mu }_{n}})\\
-3\mu ^{2}M_{2}\, G_{c_{1}}(X_{\widetilde{\mu }_{n}})\, -\, (\mu ^{3}\, +\, \mu ^{2}M_{2})\, G_{d_{1}}(X_{\widetilde{\mu }_{n}})\, \}\\
+3\times \{\mu G_{b_{3}}(X_{\widetilde{W}_{n}},X_{\widetilde{\mu }_{n}})+(M_{2}-2\mu )G_{b_{4}}(X_{\widetilde{W}_{n}},X_{\widetilde{\mu }_{n}})+(-\mu +M_{2})G_{c_{4}}(X_{\widetilde{\mu }_{n}})+(4\mu +M_{2})G_{c_{3}}(X_{\widetilde{\mu }_{n}})\\
-\, (\mu \, +\, M_{2})\, G_{d_{2}}(X_{\widetilde{\mu }_{n}})\, \}\\

\end{array} \)

\( \begin{array}{c}
\\
A_{1c_{2}}(X_{\widetilde{W}_{n}},X_{\widetilde{\mu }_{n}})\, =\, \frac{\mu M^{2}_{2}}{[M^{2}_{2}\, -\, \mu ^{2}]}[\, F_{c_{2}}(X_{\widetilde{\mu }_{n}})\, +\, F_{c_{2}}(X_{\widetilde{W}_{n}})\, ]\, +\, \frac{2\mu ^{2}M^{2}_{2}}{\overline{M}^{2}_{n}}\, F_{b}(X_{\widetilde{\mu }_{n}})\, -\, 3(2\mu \, +\, M_{2})\, G_{e}(X_{\widetilde{\mu }_{n}})\\
\\

\end{array} \)

ACKNOWLEDGMENTS. 

I wish to thank Dr.S.Lavignac and Dr.G.Moultaka for helpful comments.This work
was supported by G.A.M laboratory(http://gamweb.in2p3.fr) and LPC laboratory(http://clrwww.in2p3.fr)

\end{document}